\documentclass[fleqn,usenatbib]{mnras}

\usepackage{newtxtext,newtxmath}

\usepackage[T1]{fontenc}

\DeclareRobustCommand{\VAN}[3]{#2}
\let\VANthebibliography\thebibliography
\def\thebibliography{\DeclareRobustCommand{\VAN}[3]{##3}\VANthebibliography}

\usepackage{graphicx}	
\usepackage{amsmath}	
\usepackage{float}

\newcommand{\lSect}[1]{{\label{sec:#1}}}
\newcommand{\lFig}[1]{{\label{fig:#1}}}

\newcommand{\lTab}[1]{{\label{tab:#1}}}
\newcommand{\FIGFF}[2]{{\ref{fig:#2}{#1}}}

\newcommand{\FIG}[2]{{Fig.~\FIGFF{#1}{#2}}}
\newcommand{\Fig}[1]{{\FIG{}{#1}}}

\newcommand{\Sectff}[1]{{\ref{sec:#1}}}
\newcommand{\Sect}[1]{{\S\Sectff{#1}}}

\title[Rapidly rotating red giants]{Spectroscopic identification of rapidly rotating red giant stars in APOKASC-3 and APOGEE DR16}

\author[Patton et al.]{
Rachel A. Patton,$^{1,2}$\thanks{E-mail: patton.502@osu.edu}
Marc H. Pinsonneault,$^{1,2}$
Lyra Cao,$^{1,2}$
Mathieu Vrard,$^{1,2}$
Savita Mathur,$^{3,4}$
\newauthor
Rafael A. Garc\'{i}a,$^5$
Jamie Tayar,$^6$
Christine Mazzola Daher,$^2$
and Paul G. Beck$^{3,4,7}$
\\
$^{1}$Department of Astronomy, The Ohio State University, 140 West 18th Ave, Columbus, OH 43210, USA\\
$^{2}$Center for Cosmology and AstroParticle Physics, The Ohio State University,
191 West Woodruff Avenue, Columbus, OH 43210\\
$^{3}$Instituto de Astrofísica de Canarias (IAC), 38205 La Laguna, Tenerife, Spain\\
$^4$Universidad de La Laguna (ULL), Departamento de Astrofísica, 38206 La Laguna, Tenerife, Spain\\
$^5$AIM, CEA, CNRS, Université Paris-Saclay, Université de Paris, Sorbonne Paris Cité, 91191 Gif-sur-Yvette, France\\
$^6$Department of Astronomy, University of Florida, Bryant Space Science Center, Stadium Road, Gainesville, FL 32611, USA\\
$^7$ Institut für Physik, Karl-Franzens Universität Graz, Universitätsplatz 5/II, NAWI Graz, 8010 Graz, Austria
}

\date{Accepted XXX. Received YYY; in original form ZZZ}

\pubyear{2023}

\begin{document}
\label{firstpage}
\pagerange{\pageref{firstpage}--\pageref{lastpage}}
\maketitle

\begin{abstract}
Rapidly rotating red giant stars are astrophysically interesting but rare. In this paper we present a catalog of 3217 active red giant candidates in the APOGEE DR16 survey. We use a control sample in the well-studied \textit{Kepler} fields to demonstrate a strong relationship between rotation and anomalies in the spectroscopic solution relative to typical giants. Stars in the full survey with similar solutions are identified as candidates. We use \textit{v}sin\textit{i} measurements to confirm 50+/- 1.2 \% of our candidates as definite rapid rotators, compared to 4.9+/-0.2\% in the \textit{Kepler} control sample. In both the \textit{Kepler} control sample and a control sample from DR16, we find that there are 3-4 times as many giants rotating with 5 < \textit{v}sin\textit{i} < 10 km s$^{-1}$ compared to \textit{v}sin\textit{i} > 10 km s$^{-1}$, the traditional threshold for anomalous rotation for red giants. The vast majority of intermediate rotators are not spectroscopically anomalous. We use binary diagnostics from APOGEE and \textit{Gaia} to infer a binary fraction of 73+/-2.4 \%. We identify a significant bias in the reported metallicity for candidates with complete spectroscopic solutions, with median offsets of 0.37 dex in [M/H] from a control sample. As such, up to 10\% of stars with reported [M/H]<-1 are not truly metal poor. Finally, we use Gaia data to identify a sub-population of main sequence photometric binaries erroneously classified as giants.
\end{abstract}

\begin{keywords}
stars: rotation -- stars: binaries: general -- stars: low-mass
\end{keywords}



\section{Introduction}
Rapid rotation in low-mass giant stars is a sign of a star with an unusual history. For stars born below the Kraft break \citep[T$_\mathrm{eff}$ < 6250 K; M < 1.3 M$_\odot$][]{Kra67}, the convective envelope is deep enough that magnetized winds form, which carry away angular momentum. These low mass stars rotate slowly, even on the main sequence. Even higher mass stars that leave the main sequence rotating more rapidly experience magnetic braking when they become giants with deep surface convection. Furthermore, all giants expand and slow down through angular momentum conservation. The net effect is that the bulk of red giants are very slow rotators, typically with undetected rotation rates \citep[][]{deM96}.

Yet we know rapidly rotating red giants exist, as there are several ways in which rapid rotation manifests itself. Rotational broadening of spectral lines corresponding to \textit{v}sin\textit{i} > 10 km s$^{-1}$ is the typical threshold adopted in the literature to designate rapid rotation in red giants \citep[e.g.][]{Tay15}. With the advent of large time-domain surveys, star spot modulation can also be used to infer rotation periods for giants \citep[e.g.,][]{Gar14c}. Magnetic activity, caused by a rotationally-driven, convective dynamo in the giant's envelope results in star spots \citep[e.g.,][]{Yad15}, thus rotational modulation in its light curve \citep[e.g.,][]{Mat14}. Position in an HR diagram can also indicate binary interaction and current rapid rotation. For example, there is a field turnoff related to the finite age of the Galaxy, and a corresponding minimum effective temperature in the red giant branch. Stars cooler than this locus, sometimes referred to as sub-subgiants, are preferentially active stars, and frequently interacting binaries \citep{Gel17a,Lei17,Gel17b,Lei21b}. Stars with typical evolutionary histories are either completely absent or nearly so in these domains. 

Still, each of these detection methods have their shortfalls. In principle, rotation in the form of \textit{v}sin\textit{i} can be measured directly from spectra. However, the large majority of giants are slow rotators, and adding rotation as an additional degree of freedom in spectroscopic fitting pipelines is computationally expensive for a quantity that is an upper limit for 98\% or more of the sample. Furthermore, rapidly rotating giants have spectra that can be difficult to fit with conventional methods. Time-domain surveys are biased towards higher amplitude spot modulation and can miss lower amplitude modulation from targets with moderate rotational enhancement. Although detected from rotational line broadening, \citet{Tay15} found a significant population of moderately rotating (5 < \textit{v}sin\textit{i} < 10 km s$^{-1}$) red clump stars. Such low-amplitude rotators should exist but are likely to be missed in photometric surveys. Lastly, it is not guaranteed that all or most rapidly rotating red giants are photometrically distinct from their non-rotating counterparts. New diagnostics are needed to extract the underlying population of rapidly rotating red giants.

As such, the census of rapidly rotating red giants is largely incomplete, and their expected occurrence rates remain uncertain due to the broad range of values from observations. \citet{Mas08} found a 0.3\% rotation fraction from 761 Hipparcos giants. \citet{Car11} found that 2.2\% of their sample of roughly 1300 K giants was rapidly rotating. \citet{Dah22} found that 0.8 - 3.5\%, depending on the threshold, of 79 308 field giants rapidly rotate. In the \textit{Kepler} field, where the stars are well characterized, \citet{Tay15} identified 10 rapidly rotating stars (0.5\%) in a sample of 1950 red giants with asteroseismic detections in the APOKASC-1 survey \citep{Pin14}, \citet{Cei17} measured surface rotation periods for 361 (2\%) rapidly rotating and active red giants from a sample of 17 377 red giants, and, most recently, \citet{Gau20} identified 370 (out of 4500; 8\%) red giants, some single, some in binaries, displaying rotational modulation in their light curves. Selection effects in samples, differences in the definition of what constitutes a rapid rotator, and true differences in distinct stellar populations, likely account for this dispersion. Furthermore, the bulk of the above studies on rapidly rotating red giants have been done on targets in the \textit{Kepler} field, which we cannot assume is a representative sample of the entire Galaxy. 

Knowing the true underlying population of rapidly rotating red giants is desirable for a variety of reasons. First, binary interaction is the dominant cause of rapid rotation in red giants. Main sequence mergers can produce blue stragglers, which tend to rotate more rapidly than old field stars \citep[see, for example,][for blue stragglers in M67]{Pet84}. Subgiant mergers with stellar or planetary companions will also produce fast spinning stars \citep[e.g.][]{Lei17}. Both are a natural source for rapidly rotating single giants. On the lower giant branch, the timescale for tidal synchronization is shorter than the evolutionary time scale, so close binaries will be tidally synchronized \citep{Zah77,Ver95}. This produces much more rapid rotation than expected from single star expansion and angular momentum conservation. Even on the main sequence, tidal synchronization with a binary companion is a significant contributor to rapid rotation \citep{Zah89}. Not only are these interesting systems in their own right, enabling the study of, for example, synchronization timescales and formation channels of white dwarf-white dwarf binaries/supernova type Ia progenitors, but they also allow us to calibrate uncertain binary interaction and population synthesis models, such as merger rates and the physics of mass transfer \citep[e.g.][]{Lei21}. Additionally, a large sample of rapidly rotating giants, the majority of which we expect to be in binaries, allows us to probe activity metrics in binaries, the demographics of active giants \citep[e.g.,][]{Gau20}, and the types of evolutionary tracks which populate non-standard parts of the HR diagram \citep[e.g.][]{Lei21b}.

The trouble lies in finding large samples of rapidly rotating, red giants. Typical studies of rapidly rotating red giants only yield tens to hundreds of stars. This makes sense given the initial sample sizes. But in larger stellar samples, on the order of hundreds of thousands of stars, we want to be able to easily identify the thousands of expected rapidly rotating red giants without searching for rotation signatures giant by giant. Furthermore, rapid rotation, and the activity caused by it, is known to degrade data quality. It can suppress asteroseismic signals \citep[e.g.,][]{Gar10,Tay15,Mat19,Gau20}, add extra modulation to light curves \citep[e.g.,][]{Mat14}, and prevent good fits to spectra due to the dramatically different temperatures of spotted and non-spotted surfaces \citep{Cao22}. Rapid rotators produce data that may not survive quality cuts, leading to an underestimate of their true numbers. To get a true representative sample from a large stellar population, we need to be able to find rapid rotators even among stars with poor quality data. 

Data Release 16 (DR16) from the Apache Point Observatory Galactic Evolution Experiment \citep[APOGEE;][]{Maj17,Wil19,Ahu20} contains high-resolution (R of 22,500) infrared spectra for over 437,445 nearby stars. This survey is ideal for locating a large data set of active, evolved stars. The manner in which stellar spectra are fit in this survey results in natural bins by which we can sort stars and search for spectroscopic anomalies indicating rotation. 

In this study, we explore whether we can use anomalies in spectroscopic solutions to preferentially identify rapidly rotating giants. We use the well-studied \textit{Kepler} field to infer a census of rapid rotators, use asteroseismic data to distinguish between different evolutionary states, and demonstrate that our technique recovers the majority of these targets in that sample. We describe the catalogs from which we draw our data in section \Sect{dat}. We then evaluate the efficacy of each spectroscopic criterion at identifying rapid rotators in \Sect{criteria}. Having defined new spectroscopic criteria to quickly identify rapidly rotating red giants, we construct a DR16 catalog of several thousand rapidly rotating red giant candidates and validate them with \textit{v}sin\textit{i} measurements, which we discuss in section \Sect{dr16}. \Sect{disc} discusses the implications of a large population of rapidly rotating red giants, biases in our selection criteria, and how adaptable the criteria are for other data releases and other spectroscopic surveys. Finally, we summarize our results in \Sect{con}.

\section{Data}
\lSect{dat}

For photometric and spectroscopic identification of red giants as well as independent confirmation of rotation, we drew data from several sources. We describe each catalog and its utility below.

\subsection{APOGEE DR16}
APOGEE uses an automated pipeline to infer stellar parameters from a target's spectra. The APOGEE Stellar Parameter and Chemical Abundances Pipeline \citep[ASPCAP;][]{Nid15, Gar16} extracts stellar parameters in two parts. First, the code FERRE \citep{All06} matches the input spectrum to the best-fit template from a grid of dwarf and giant synthetic spectra, by minimizing $\chi^{2}$ and returning estimates for stellar parameters. This is an 8-D fit to get log $g$, $T_{\mathrm{eff}}$, carbon abundance, nitrogen abundance, overall metal abundance, $\alpha$-element abundance, microturbulent velocity, and either \textit{v}sin\textit{i} or macroturbulent velocity for dwarfs and giants respectively. In the former case, mactroturbulent velocity is assumed to be negligible and in the latter, \textit{v}sin\textit{i} is fixed at 1.5 km s$^{-1}$.

Synthetic spectral grids cover G-M type giants and F-M dwarfs, with the giant grids spanning -0.5< log $g$ < 4.5 cm s$^{-2}$, depending on spectral type, and the dwarf grids spanning 2.5 < log \textit{g} < 5.5 cm s$^{-2}$ \citep{Jon20}. Stars can be fit using multiple grids, where the model yielding the lowest $\chi ^2$ gets adopted. Due to the overlap in log \textit{g} between the dwarf and giant grids, some stars will be fit with both giant and dwarf template spectra. In these instances, the goodness-of-fit is not only important for determining stellar parameters but also evolutionary state. Spectra that cannot be fit at this stage trigger the appropriate quality control flag and the pipeline returns no solution.

\begin{figure}
    \centering
    \includegraphics[scale=0.4,trim={0.7cm 7.5cm 0.1cm 7.0cm},clip]{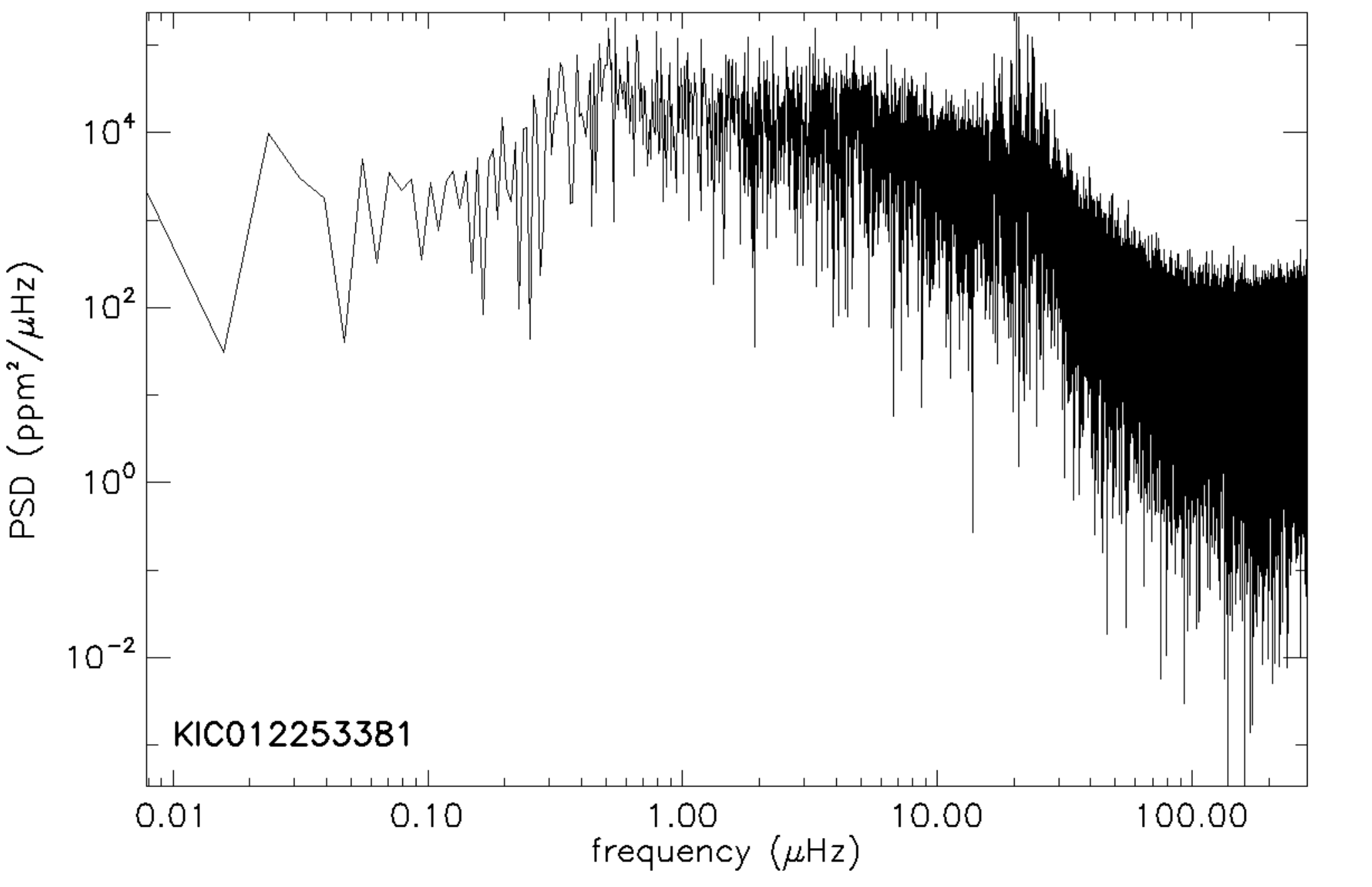}
    \includegraphics[scale=0.42,trim={1.4cm 8.3cm 0.0cm 7.0cm},clip]{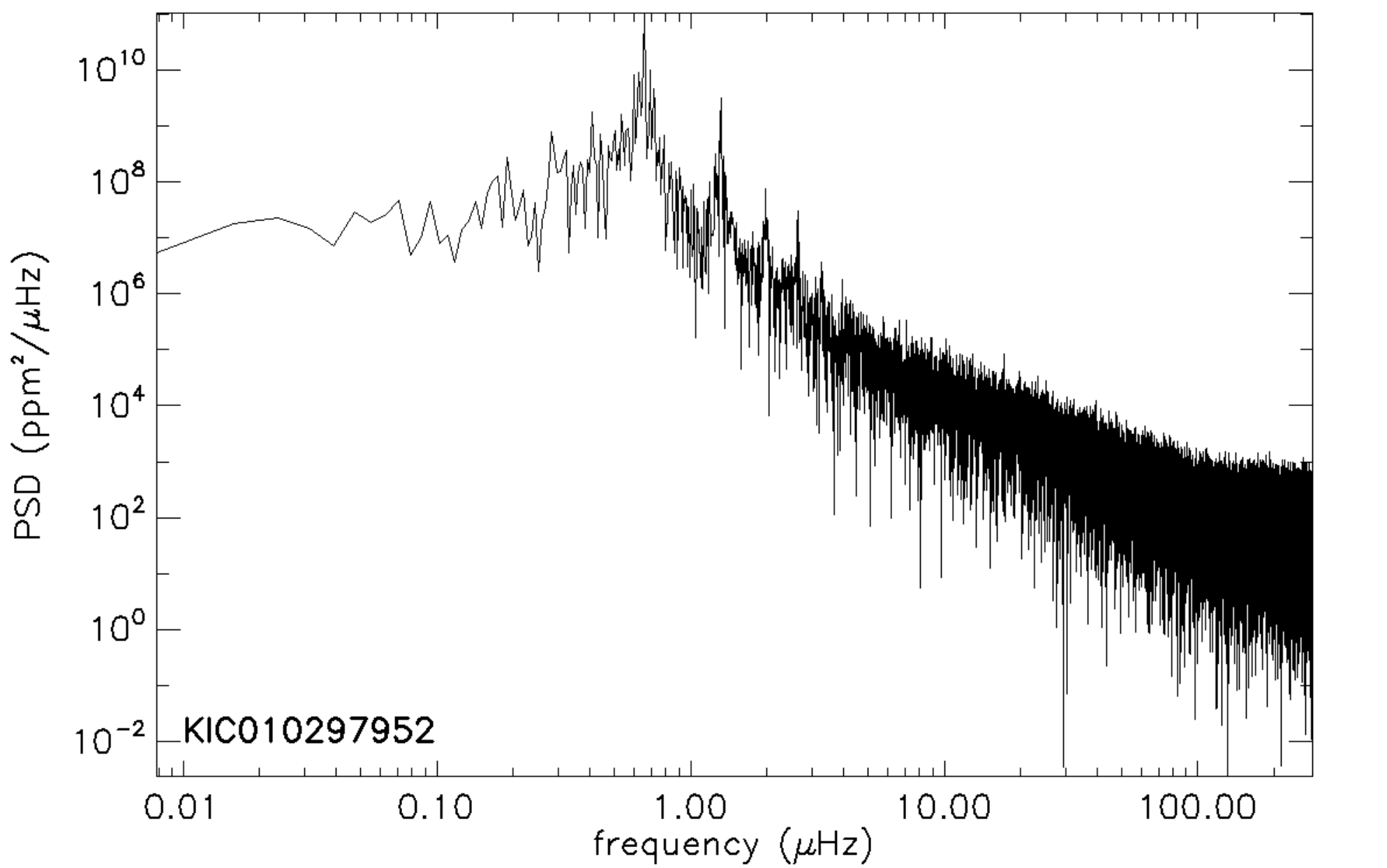}
    \caption{Power spectra for two stars that have complete spectroscopic solutions but are anomalously cool. The top panel shows the power spectrum for KIC12253381, a confirmed slowly rotating giant with solar-like oscillations. Note the characteristic cluster of oscillation modes centered near 20 $\mu$Hz. The bottom panel shows the power spectrum for KIC10929752, a confirmed rapidly rotating giant with a strong rotation signature around 0.5 $\mu$Hz and its harmonics around 1.5 and 2 $\mu$Hz.}
    \lFig{PSD}
\end{figure}

Log $g$ and $T_{\mathrm{eff}}$ estimates from the FERRE fit are then calibrated against independent measures of the same parameters. For giants, the calibrations come from asteroseismology for log $g$ and trends in photometric estimates for $T_{\mathrm{eff}}$ for stars with minimal reddening. Spectra without the calibrated stellar parameters we dub partial solutions, whereas spectra that have both the initial FERRE fits and the calibrated stellar parameters we deem complete solutions. 

We might expect rapid rotators to preferentially fall into the first two categories because rapid rotation can cause a non-uniform surface temperature, especially in active, cool stars with large spot filling factors, making a single-temperature solution difficult \citep{Cao22}. However, rapid rotation still affects data quality even among stars with complete solutions. ASPCAP does not include rotation as a free parameter when fitting the spectra of giants. The derived stellar parameters are then strongly perturbed by excess line broadening, which can be artificially produced by correlated changes in effective temperature, surface gravity, and elemental abundances. Rapidly rotating giants whose spectra are fit without rotation as a free parameter have artificially high surface gravity \citep{Tho19} and artificially low effective temperatures \citep{Dix20}.

Among the stars with complete solutions, we still need a tool to preferentially identify rapidly rotating red giants. To construct this tool, two pieces are essential: a definition for characteristic properties of typical red giants and a method for identifying outliers. Conveniently, DR16 gives us the definition of a typical red giant due to the inclusion of a means to spectroscopically separate first-ascent red giant branch (RGB) stars from red clump (RC) stars. This is possible because there are systematic surface gravity and temperature offsets between the two populations. 

 From stellar theory, we expect there to be a mapping between mass, metallicity, $\alpha$-enrichment, and the location of the RGB on an HR diagram. DR16 defined the mean locus of the RGB using a reference temperature, $T_\mathrm{ref}$, which represents the expected effective temperature for an average red giant of a given mass and metallicity \citep{Jon20}. The carbon-to-nitrogen ratio can be used as a proxy for mass. \citet{Hol18} found this reference temperature by fitting the average $T_\mathrm{eff}$ of red giants in the APOKASC-2 catalog \citep{Pin18}. The difference between the star's uncalibrated spectroscopic $T_\mathrm{eff}$ and $T_\mathrm{ref}$ is then used to calculate a reference number based on the temperature difference and a star's total, uncalibrated metallicity [M/H]. If a star's uncalibrated carbon-to-nitrogen ratio [C/N] is less than the reference number, then it is classified as an RGB star. 

Combining the [C/N] dependence with the reference temperature, we can write the condition for separating RGB stars from RC stars as 
\begin{equation}
   \Delta T = T_\mathrm{eff, spec} - T^{'}_\mathrm{ref}
\end{equation}
where $T_\mathrm{eff, spec}$ is the uncalibrated spectroscopic effective temperature and 
\begin{equation}
    T^{'}_\mathrm{ref} = 3032.786 - 357.13 \mathrm{[C/N]} + 552.6 \mathrm{log} g - 488.886 \mathrm{[M/H]}
\end{equation}
for uncalibrated values of log $g$, [C/N], and [M/H]. If $\Delta T$ is negative, the star is on the RGB. This evolutionary state separation only is calibrated to the lower giant branch, for 2.38 < log $g$ < 3.5 cm s$^2$. However, to capture the sub-subgiants, which are interacting red giants, we must extend the range of log \textit{g} to 3.85 cm s$^2$. We do not have to worry about contamination from true subgiants, however, because they tend to be hotter and will not scatter into the sub-subgiant domain.

To understand the origin of the rapid rotation we must assess binarity. Radial velocity (RV) variations suggest the presence of a companion. Since we expect most of the rapid rotators to be close, interacting binaries, the RV amplitudes should be high, so we can adopt a more lax criterion, requiring at least two visits instead of more visits. We define as binaries those stars which have a scatter around the average heliocentric velocity, \texttt{VSCATTER}, greater than 3*\texttt{VERR\_MED}, the average measurement error of the heliocentric velocity. This criterion does not exclusively identify close binaries, however. We deem close binaries those stars with at least two APOGEE visits where \texttt{VSCATTER} > 10*\texttt{VERR\_MED}. This choice is comparable to a $\Delta RV_\mathrm{max}$ of 1 km s$^{-1}$, the threshold adopted in \citet{Moe19} and \citet{Dah22}.

Lastly, ASPCAP provides a lengthy list of quality control flags to indicate any issues with the fit. in the instance of rapid rotation, ASPCAP will return \texttt{SUSPECT\_BROAD\_LINES},\texttt{ROTATION\_WARN}, or \texttt{ROTATION\_BAD}. We include these flags in our final catalog of rapid rotators. 

\subsection{APOKASC-3 and rapidly rotating red giant catalogs from \textit{Kepler}}

We need a control sample containing well-studied, well-characterized stars in which we can test various rotation diagnostics. Any of the three criteria which is particularly good at identifying rapid rotators in the control sample can then be applied to the entire DR16 catalog, picking out stars which are most likely to be rapid rotators instead of searching for rotation giant by giant. 

The APOKASC-3 catalog (Pinsonneault et al., in prep) is an ideal sample to use as our control because it combines the IR spectroscopy of APOGEE with high cadence photometry from \textit{Kepler} for 23 363 stars observed by both surveys. For our purposes, the key advantages of this over other public catalogs \citep{Pin18, Yu18} is that the full set of giants has been studied for both seismology and rotation, and all of the giants in the \textit{Kepler} fields targeted by APOGEE are included. The APOGEE survey selected giants in the field based on color and magnitude cuts, and as a result the sample is not biased toward asteroseismic detections. This is important, because rapid rotators are selected against in purely asteroseismic samples \citep{Tay15,Gau20}. 

Giants in APOKASC-3 are selected for having log \textit{g} < 3.6 cm s$^{-2}$ and $T_\mathrm{eff}$ < 6000 K. We verify the spectroscopic identification of giants with \textit{Gaia} photometry. Additionally, for any target which did not have a complete solution, giants can only be selected from \textit{Gaia} photometry. Giants are defined as having a dereddened \textit{Gaia} B$_\mathrm{P}$ - R$_\mathrm{P}$ color > 0.9 and dereddened, absolute \textit{Gaia} G-band magnitude < 0.3 + 3.4 * \textit{Gaia} color. This is a strict photometric cut and it does remove the blue end of the clump, but we are not losing many targets. In total, there are 15 220 giants. 

We use these giants as our control sample, testing the efficacy of each spectroscopic criterion at picking out rapid rotators. To do this, we need a uniform measure of rapid rotation. In short, we measure \textit{v}sin\textit{i} for all giants in APOKASC-3, the details of which are in \Sect{Apo3_rot}. The method to measure \textit{v}sin\textit{i} we use only works for stars with a best-fit template spectrum, meaning stars with no spectroscopic solution are not included in the sample of rapid rotators. In this instance, we crossmatch the 32 giants with no solution to various catalogs of rapidly rotating red giants in \textit{Kepler} as well as \textit{Gaia}. We check the \citet{Gau20} binary catalog and extract orbital periods and/or rotation velocities at the target's equator, where measured, for stars which appear in that catalog. If a star only has a measured rotation period and no other indicators of rotation, we calculate rotation speed using the radii from \citet{Ber20}. Any speed > 10 km s$^{-1}$ is considered rapid. Then we check the \citet{Cei17} catalog of red giants with measured surface rotation periods and extract a rotation period for the one star, KIC 7531136, in our sample which appears in this catalog. From this rotation period and a radius estimate \citep{Ber20}, we infer a rotation speed of 16 km/s. Stars which appeared in any of these catalogs, and had measured orbital periods or v sin{i} > 10 km s$^{-1}$ where applicable, were designated as rapid rotators. We maintain the 10 km s$^{-1}$ threshold for these measurements of equatorial rotation speed to be conservative. 

We also inspect their light curves. Using the \textit{Kepler} KEPSEISMIC\footnote{https://archive.stsci.edu/prepds/kepseismic/} lighcurves and power spectrum densities calibrated by the KADACS pipeline \citep{Gar11, Gar14b}, we inspect the stars' power spectra for rotation signatures. While the presence of a high amplitude peak at low frequency along with its harmonics in some cases can provide an estimate of the surface rotation period of a star \citep[e.g., ][]{San19, San21}, the modulation can come from a nearby star polluting the \textit{Kepler} light curve as also explained in \citet{Cei17} and \citet{Col17}. The cross-check with the APOGEE data should allow us to discard those cases. If the analysis from  both instruments yield a fast rotation, we can consider that this star has a high surface rotation rate. However we should be cautious that cases of pollution by a nearby star can occur in \textit{Kepler} data. An example of a high-amplitude, low-frequency peak is shown in the lower panel of \Fig{PSD} compared to a giant exhibiting normal oscillation modes shown in the top panel. Any star which displays a rotation peak in its power spectrum we deem a rapid rotator. 

Finally, we check SIMBAD for signs of variability and any other information on the nature of the star. Stars which were labeled as rotational or ellipsoidal variables, as well as eclipsing binaries, were designated as rapid rotators.

\subsection{\textit{Gaia}}
For stars without complete spectral solutions and as an additional check for stars with complete spectroscopic solutions, we photometrically identified the giants using \textit{Gaia}. All stars in our sample are found in \textit{Gaia} Data Release 2 (DR2) \citep{Gai16,Gai18b}. We choose to use DR2 to remain consistent with the \citet{Bai18} distance estimates. Since the photometry is used only for a coarse sorting of evolutionary states, APOGEE giants are bright and 2.9 Kpc away on average, so zero-point issues are not as large, and DR3 results agree well on average with DR2, we find DR2 to be sufficient. We de-redden our stars using the same methods as \citet{God21} based on the process described in \citet{Gai18a}, where extinction coefficients depend on both the \textit{Gaia} B$_\mathrm{P}$ - R$_\mathrm{P}$ color and A$_\mathrm{V}$. We obtain $A_V$ for both the APOKASC-3 and for the entire APOGEE DR16 sample by integrating over 3D dust maps using the Python package \texttt{mwdust} \citep{Bov16}. We use the combined dust maps of \citet{Mar06} and \citet{Gre19}. When converting from $A_V$ to \textit{Gaia}-band extinctions, we do not adopt any of the singular values from the literature because the conversion factor for each band depends on the temperature of the star. Instead, we iteratively solve for the conversion factors following the methods in \citet{God21}. Once stars are placed on the \textit{Gaia} color-magnitude diagram (CMD), we then make a cut on the upper bound of the main sequence, separating the dwarfs below from the red giants above. The exact slope of this line is not important; the line simply must exclude the main sequence. As such, the cuts made for the APOKASC-3 data differ from those made for the APOGEE data due to the larger color range spanned by the main sequence in APOGEE. Examples of our cuts are shown in \Fig{partial} and \Fig{dr16_part}.

With the recently released spectra from \textit{Gaia} DR3 we can also verify rotation and binarity for a subset of APOKASC-3 and DR16 giants bright enough to be in the sample. The parameter $v_\mathrm{Broad}$ captures the total line broadening due to rotation (v sin$i$), macroturbulence (v$_\mathrm{macro}$), and other sources. Macroturbulence in a cool giant is on the order of 2 - 3.5 km s$^{-1}$, so in principle, in the instance of a rapid rotator, rotation is the dominant source of broadening \citep{Doy14,Fre22}. If we treat v$\mathrm{Broad}$ as a proxy for \textit{v}sin\textit{i}, however, we significantly overestimate the number of rapid rotators. At a 10 km s$^{-1}$ threshold in $v_\mathrm{Broad}$, which \citet{Fre22} claims is comparable to \textit{v}sin\textit{i} for cool giants with 10.5 < G$_\mathrm{RVS}$ magnitude < 11.5, we find that $\sim$ 40\% of APOKASC-3 giants and DR16 giants with the appropriate magnitude are ``rapidly rotating."  As such, we do not use $v_\mathrm{Broad}$ to draw conclusions about the rates of rapid rotation in giants. Phillips et al. (in prep) sees that there is good agreement between \textit{v}sin\textit{i} and $v_\mathrm{Broad}$ above 20 km s$^{-1}$, so we adopt this as our threshold for rapid rotation just for giants without a spectroscopic solution, where we have limited additional constraints on rapid rotation. 

\textit{Gaia} also lets us probe binarity. High values (> 1.4) of the renormalized unit weight error (RUWE) often indicate the presence of a wide binary companion, which impacts the astrometric solution \citep{Bel20,Sta21,Ker22}. Adopting a less conservative cut, we label as binaries all targets with RUWE > 1.2. DR3 also contains the non-single star catalog, which contains targets with  complete binary classifications. More broadly we can also find radial velocity variable stars using the criteria laid out in \citet{Kat22}. While binarity alone does not imply rapid rotation, binarity can suggest the origins of rotation. Binary rapid rotators are likely tidally synchronized, whereas single rapid rotators are likely mergers. 

\section{Spectroscopic selection criteria}
\lSect{criteria}
The aim of this work is to develop a set of spectroscopic criteria which could identify, with high fidelity, rapidly rotating red giants in the full APOGEE DR16 catalog. Natural spectroscopic categories to search for rapid rotators arise from the steps ASPCAP takes to process spectra. ASPCAP assesses the quality of the spectra as they travel through the pipeline, rejecting spectra outright and returning no solution, performing an initial fit and rejecting spectra whose fits do not pass quality control, and finally doing a calibrated fit to spectra which do not trigger any quality control flags. Below, we assess how well each spectroscopic category, those with no, partial, and complete solutions, preferentially picks out rapidly rotating giants. 

\subsection{Anomalously cool stars with complete spectroscopic solutions}
\lSect{cool_crit}

\begin{figure}
    \centering
    \includegraphics[scale=0.57,trim={0.1cm 0.1cm 0.5cm 0.5cm},clip]{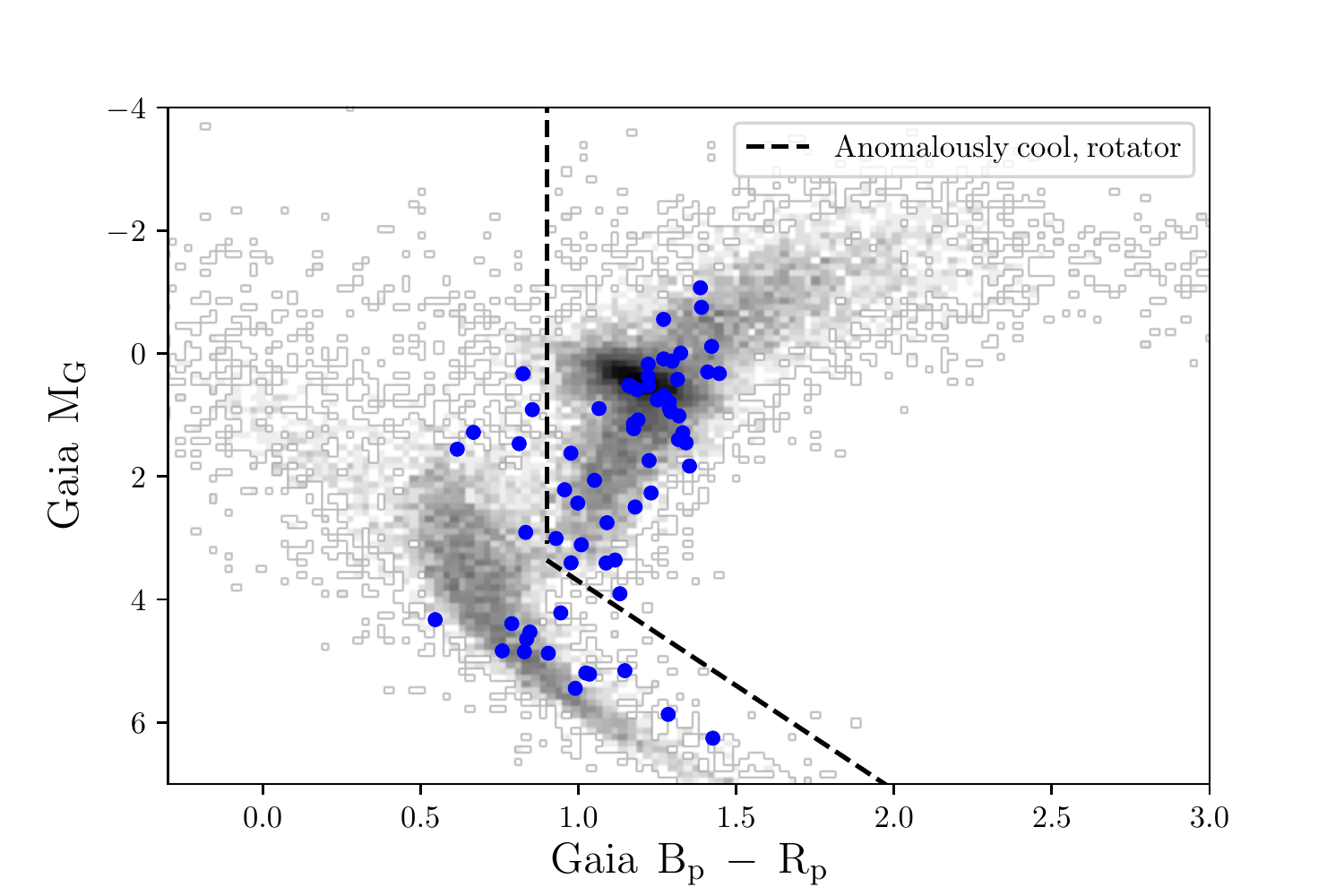}
    \includegraphics[scale=0.57,trim={0.1cm 0.1cm 0.5cm 0.5cm},clip]{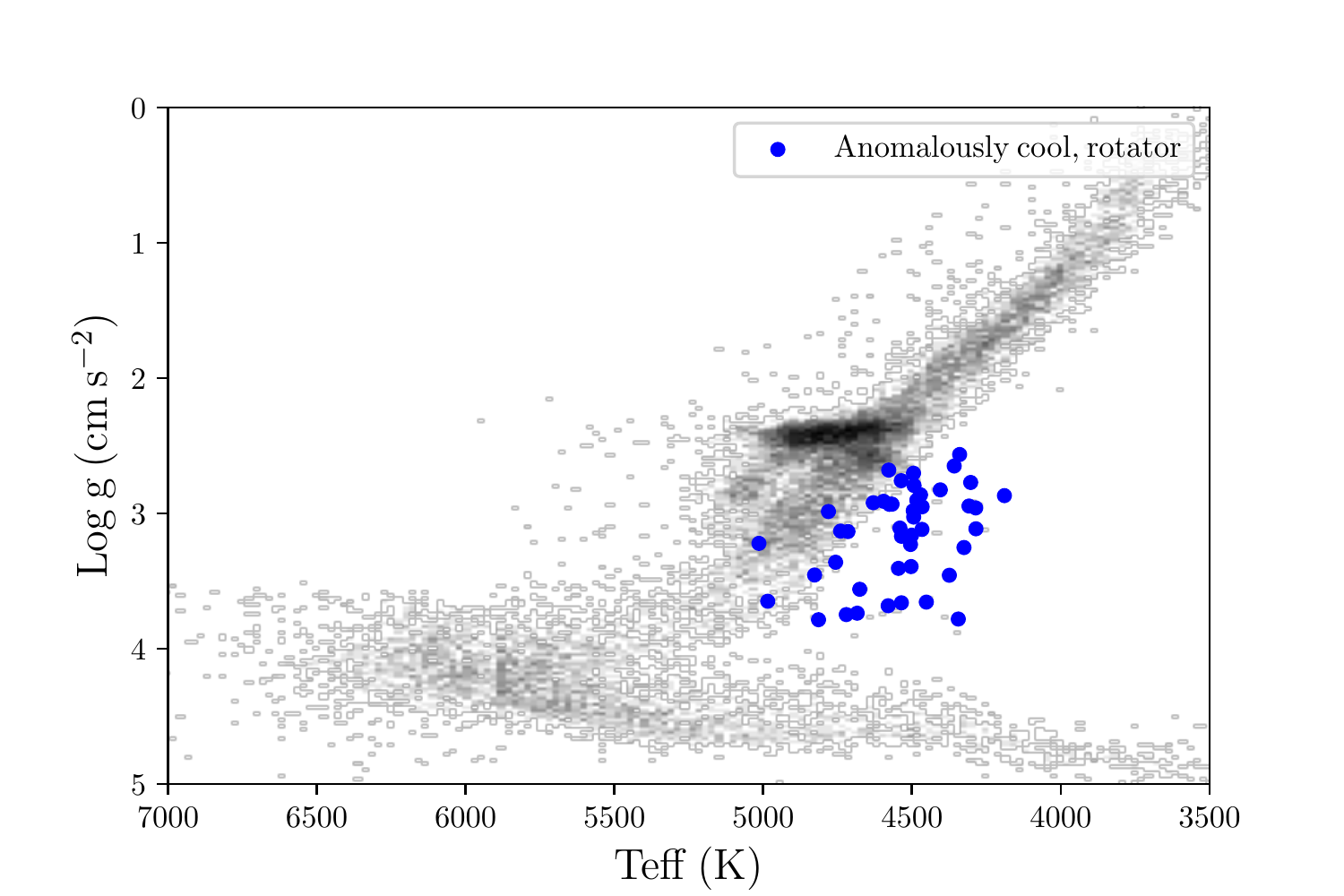}
    \includegraphics[scale=0.57,trim={0.1cm 0.1cm 0.5cm 0.5cm},clip]{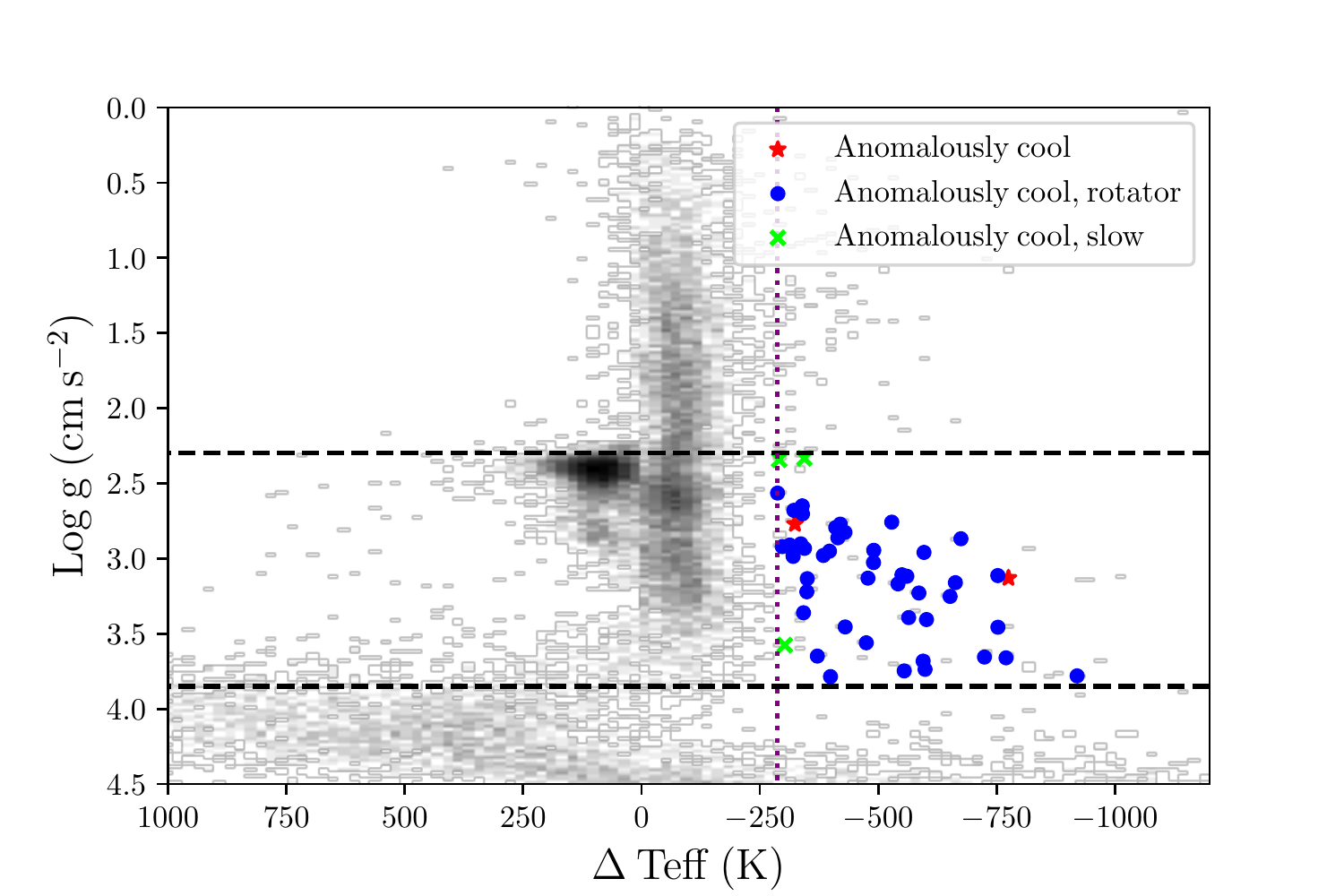}
    \caption{\textit{Top}: A \textit{Gaia} CMD of APOKASC-3 (grey) and the rapidly rotating, anomalously cool stars shows that rapid rotators cannot be distinguished from regular red giants. Photometric binaries and hot stars are excluded from subsequent plots. \textit{Middle}: Calibrated log \textit{g} versus $T_\mathrm{eff}$ for APOKASC-3 (grey) and the rapid rotators, which are more distinguishable from the RGB, but not completely. \textit{Bottom}: Calibrated log \textit{g} versus $\Delta T$ for all of APOKASC-3 (grey) and the anomalously cool sample. The $\Delta T$ calculation is only valid for stars with 2.3 < log \textit{g} < 3.5 cm s$^{-2}$, though we extrapolate to higher log \textit{g}. $\Delta T$ calculations outside this log \textit{g} range are illustrative only. Black dashed lines denote the log \textit{g} limits and the purple dotted line marks the 3$\sigma$ threshold. The anomalously cool stars are classified as rapid rotators (blue points), slow rotators (green crosses), or having not enough info (red stars).}
    \lFig{cool}
\end{figure}

The evolutionary state separator (equations 1 and 2), besides being able to isolate the RGB from the red clump, provides a novel way to quickly identify stars with stellar properties which significantly deviate from those of standard red giants. A temperature offset relative to the mean locus is a useful figure of merit. The position of a red giant in the HR diagram is predicted to be a nearly vertical line close to the Hayashi track. The zero-point is a  function of the mass and heavy element content, and the difference between clump and giant is nearly constant. As a result, correcting for these zero point differences produces two distinct temperature domains - for clump and giant - that are a useful evolutionary state diagnostic. The abundances and log g are observables in APOGEE, and the mass is correlated with [C/N]. As a result, we can measure the temperature offset using a star's [C/N] ratio, metallicity, and surface gravity. It is valid only for giants with 2.3 < log \textit{g} < 3.5 cm s$^{-2}$ \citep{Jon20} due to the narrow log \textit{g} spanned by red clump stars \citep[see discussion in, e.g.,][]{Gir16}. Because of the evidence that the log \textit{g} measurements for rapidly rotating giants are inflated \citep{Tho19}, we examine this figure of merit for the full sample down to log \textit{g} = 4.5 cm s$^{-2}$. The vast majority of anomalously cool stars have 2.3 < log \textit{g} < 3.85 cm s$^{-2}$ (bottom panel of \Fig{cool}). This higher upper limit introduces minimal contamination from dwarfs. Stars with spectroscopic temperatures at least 3$\sigma$ cooler ($\Delta T$ < -287 K) than the mean locus of red giants of equivalent mass and metallicity we deem anomalously cool. For the few stars in our sample with total metallicity [M/H] < -1, we only deem them anomalously cool if they satisfy the metallicity-dependent criterion described in \Sect{dr16}. Ultimately we identify 71 anomalously cool stars. 

\begin{table}
\centering
  \caption{Column names for the table containing our APOKASC-3 anomalously cool rapid rotator candidate sample}
  \lTab{Apo_cool}
  \begin{tabular}{lr}
  \\
  \hline
  \textit{Kepler} ID & \citet{Bro11}\\ 
  APOGEE ID & APOGEE DR16\\
  log \textit{g} (cm s$^{-2}$) & APOGEE DR16\\ 
  $T_\mathrm{eff}$ (K) & APOGEE DR16\\
  {[M/H]} & APOGEE DR16\\
  $\Delta T$ (K) & this work\\
  m$_\mathrm{G}$ & \textit{Gaia} DR2 \\
  m$_{\mathrm{B}_\mathrm{P}}$ & \textit{Gaia} DR2\\
  m$_{\mathrm{R}_\mathrm{P}}$ & \textit{Gaia} DR2\\
  $A_V$ & this work\\ 
  distance (pc) & \textit{Gaia} DR3\\
  \textit{v}sin\textit{i} (km s$^{-1}$) & this work\\
  $v_\mathrm{Broad}$ (km s$^{-1}$) & \textit{Gaia} DR2\\
  period (d) & \citet{Cei17,Gau20}\\
  rotation peak & this work\\
  RUWE & \textit{Gaia} DR3\\
  \textit{Gaia} RV variable & this work\\
  APOGEE RV variable & this work\\
  notes & this work\\
  \hline

  \end{tabular}
\end{table}

\begin{table}
    \caption{Four different measures of binary fractions of the anomalously cool giants} 
    \centering
    \begin{tabular}{lcc}
         &  \textit{v}sin\textit{i} > 10 km s$^{-1}$ & 5 < \textit{v}sin\textit{i} < 10 km s$^{-1}$\\
         \hline
       RUWE f$_\mathrm{bin}$  & 16$\pm$8\% (5/31) & 23$\pm$13\% (3/13)\\
       \textit{Gaia} RV f$_\mathrm{bin}$ & 88$\pm$18\% (23/26) & 92$\pm$27\% (12/13)\\
       APOGEE f$_\mathrm{bin}$ & 84$\pm$21\% (16/19) & 100$\pm$33\% (9/9)\\
       APOGEE f$_\mathrm{bin,close}$ & 74$\pm$20\% (14/19) & 100$\pm$33\% (9/9)\\
    \end{tabular}
\end{table}

The top panel of \Fig{cool} shows where the rapid rotators from the anomalously cool sample fall in a \textit{Gaia} CMD. The plotted stars are indistinguishable from standard red giants despite their spectroscopic anomalies. Anomalously cool stars are not actually cooler than a typical red giant, however; they are not particularly red. Their cooler spectroscopic temperatures are a byproduct of how ASPCAP handles rapidly rotating, potentially active stars. 

Thirteen of our targets that were spectroscopically classified as giants are actually dwarfs, as inferred from \textit{Gaia} astrometry and photometry (\Fig{cool}). The cause is a rare failure mode in the APOGEE pipeline, which considers both dwarf and giant solutions for a subset of the data. In some cases, both fits were poor, but the giant solution had a slightly better goodness of fit. For both the APOKASC-3 sample and the full DR16 sample we use a \textit{Gaia} CMD cut to remove such stars from our sample. There are an additional 8 stars which are too blue to be giants. We make a cut at B$_\mathrm{P}$ - R$_\mathrm{P}$ of 0.9 to remove these targets. This leaves us with 50 targets. 

While these remaining stars appear as normal red giants in a \textit{Gaia} CMD, when we examine them in log \textit{g} - $T_\mathrm{eff}$ space, we begin to see some differentiation from typical red giants. The middle panel of \Fig{cool} shows that the majority of anomalously cool stars are cooler than the red edge of the RGB.

The separation between typical red giants and rapid rotators is best seen if instead of $T_\mathrm{eff}$ we plot $\Delta T$ versus log \textit{g}. The bottom panel of \Fig{cool} shows the two distinct populations. As expected, red clump stars fall to the left of the $\Delta T$ = 0 line while the RGB stars are clustered to the right, occupying a narrow range between 0 and -200 K. Notably, the anomalously cool stars span a range of roughly 800 K.

The majority of anomalously cool giants are rapidly rotating. Of 50 anomalously cool stars, 45 of them (90\%) are confirmed rapid rotators, with 31 having \textit{v}sin\textit{i} > 10 km s$^{-1}$ and 14 having 5 < \textit{v}sin\textit{i} < 10 km s$^{-1}$. Only 4 stars (8\%) can be confirmed as true slow rotators due to their low \textit{v}sin\textit{i} measurements. The remaining 1 star (2\%) does not have enough information to classify it as true slow rotators or rapid rotators. All data used to select and verify rotation in these targets are compiled in a table released with this paper. Table 1 lists the column names.

Table 2 lists the binary fraction of this sample from four separate measures. We expect the majority of these targets to be tidally synchronized binaries. While just about 75\% of the most rapidly rotating giants do have a close binary companion, the Poisson errors are large. For the intermediate rotators, the close binary fraction is 100\%, which is surprising since the tidally synchronized systems should have the largest spinup. However, the Poisson errors are large and the close binary fraction for rapid and intermediate rotators agrees within 1$\sigma$. It is possible that the inclination of the rapidly rotating systems is low. We also see that about 10\% of giants have a wide binary companion, as determined from \textit{Gaia} RUWE. Even when the RUWE binary detection limit is set at 1.4, the wide binary fraction is still non-zero. This is surprising since a companion at a distance cannot spin up a star. While most of the targets caught by this criterion are binaries, we see that they are not exclusively close, tidally synchronized systems like we expected.

\subsection{Stars with partial spectroscopic solutions}
\lSect{fit_no_cor}

\begin{table}
\centering
  \caption{Column names for the table containing our APOKASC-3 rapid rotator candidate sample of giants with partial or no spectroscopic solutions. Binarity is not assessed for giants with no spectroscopic solution.}
  \lTab{Apo_other}
  \begin{tabular}{lr}
  \\
  \hline
  \textit{Kepler} ID & \citet{Bro11}\\ 
  APOGEE ID & APOGEE DR16\\
  m$_\mathrm{G}$ & \textit{Gaia} DR2\\
  m$_{\mathrm{B}_\mathrm{P}}$ & \textit{Gaia} DR2\\
  m$_{\mathrm{R}_\mathrm{P}}$ & \textit{Gaia} DR2\\
  $A_V$ & this work\\ 
  distance (pc) & \textit{Gaia} DR2\\
  \textit{v}sin\textit{i} (km s$^{-1}$) & this work\\
  $v_\mathrm{Broad}$ (km s$^{-1}$) & \textit{Gaia} DR3\\
  period (d) & \citet{Cei17,Ber20}\\
  rotation peak & this work\\
  RUWE & \textit{Gaia} DR3\\
  \textit{Gaia} RV variable & this work\\
  APOGEE RV variable & this work\\
  Flags & APOGEE DR16\\
  notes & this work\\
  no solution & APOGEE DR16\\
  \hline
  \end{tabular}
\end{table}

\begin{figure}
    \centering
    \includegraphics[scale=0.57,trim={0.1cm 0.1cm 0.5cm 0.5cm},clip]{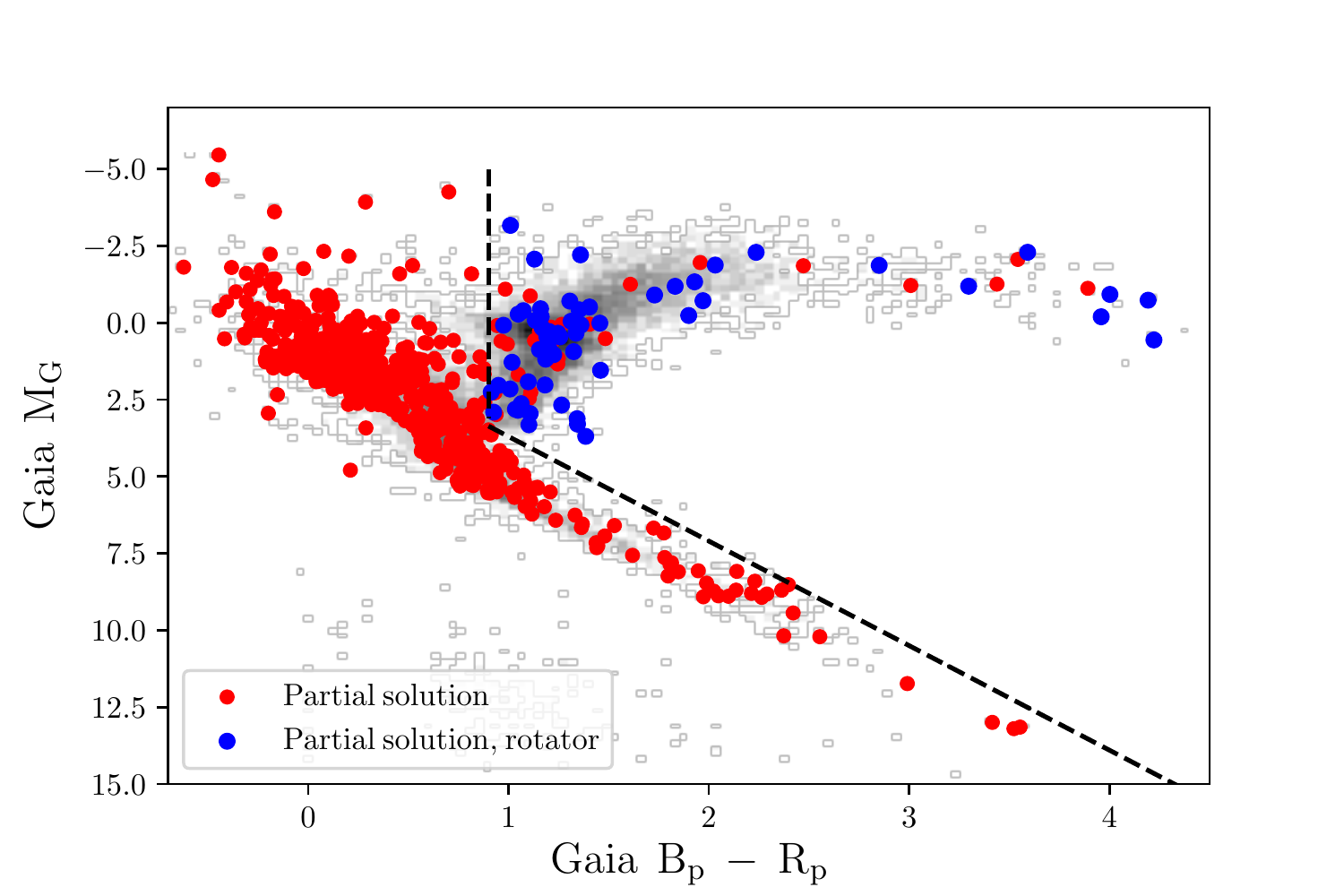}
    \caption{A \textit{Gaia} CMD of stars with partial spectroscopic solutions compared to the entire APOKASC-3 sample (grey). All targets which fall above and to the right of the dashed line we consider giants. Blue points represent confirmed rapid rotators, whereas the red stars are all stars that have partial solutions.}
    \lFig{partial}
\end{figure}

\begin{table}
    \caption{Four different measures of binary fractions of the giants with partial spectroscopic solutions.} 
    \centering
    \begin{tabular}{lcc}
         &  \textit{v}sin\textit{i} > 10 km s$^{-1}$ & 5 < \textit{v}sin\textit{i} < 10 km s$^{-1}$\\
         \hline
       RUWE f$_\mathrm{bin}$  & 27$\pm$7\% (14/52) & 27$\pm$16\% (3/11)\\
       \textit{Gaia} RV f$_\mathrm{bin}$ & 70$\pm$15\% (23/33) & 25$\pm$18\% (2/8)\\
       APOGEE f$_\mathrm{bin}$ & 63$\pm$15\% (17/27) & 71$\pm$31\% (5/7)\\
       APOGEE f$_\mathrm{bin,close}$ & 59$\pm$15\% (16/27) & 57$\pm$29\% (4/7)\\
    \end{tabular}
\end{table}

These spectra were high enough quality that the FERRE code could generate an initial fit but not high enough quality to support the subsequent calibrated line fitting. Quality is assessed in several ways. Stars with too low signal-to-noise (< 50), too high of a $\chi^2$ ($\chi^2$/(signal-to-noise/100)$^2$ > 50), too much variation between photometric and spectroscopic temperatures (> 1000 K discrepancy) from the FERRE fit, or with spectral lines that are more than twice the width of the template spectrum due to rotational broadening or line blending from double-lined spectroscopic binaries, all get flagged as bad.

Rotation can affect these spectra in two ways, first causing line broadening captured by the rotation flag mentioned above. Second, it can trigger the color-temperature flag either due to activity-induced temperature gradients on a star's surface, which are difficult to capture with a single spectroscopic temperature, or due to the presence of a binary companion, where, again, a single temperature cannot accurately reflect the combined light of the binary. While rotation is not the sole cause of poor fits, we want to quantify how significant of a role it plays in triggering the quality flags. 

We identify 684 stars in APOKASC-3 with partial spectroscopic solutions. Though we have an initial guess at the stars' spectral parameters, we choose to identify red giants by their position on a \textit{Gaia} CMD for greater accuracy. \Fig{partial} shows the CMD of this sample compared to the entire APOKASC-3 catalog. The vast majority of these stars are dwarfs, so we impose a cut at a \textit{Gaia} B$_\mathrm{P}$ - R$_\mathrm{P}$ of 0.9 mag ($T_\mathrm{eff} \approx 5450$ K) to eliminate hot stars and a cut near the base of the RGB to eliminate dwarfs and subgiants. All stars which fall above the dividing line are selected as red giants, a population of 95 in total. Table 3 lists the properties of these stars included in the table released with this paper. Note that these stars are not photometric outliers, instead lying well within the red giant branch.

Of our 95 giants, 63 (66\%) are rapid rotators, with 52 having \textit{v}sin\textit{i} > 10 km s$^{-1}$ and 11 with 5 < \textit{v}sin\textit{i} < 10 km s$^{-1}$. Some of the remaining giants are indeed rapid rotators - they have triggered the \texttt{ROTATION\_BAD} flag - but we do not measure a \textit{v}sin\textit{i}, so they do not contribute to the count. Though this criterion requires more work in separating out the red giants, the high fraction of rapid rotators it identifies, even greater than the anomalously cool stars, makes it a good tool to target rapidly rotating stars. 

The binary fraction of this sample is listed in Table 4. We did not expect a single binary configuration to explain the rapid rotation of these systems like we did for the anomalously cool giants. Giants with partial solutions span the RGB and clump and a star's position on the giant branch changes its likelihood of being in a close, tidally synchronized system. The higher wide binary fraction is surprising, though giants at the tip of the giant branch can interact with stars at a larger separation. Even when the RUWE binary detection limit is set at 1.4, the wide binary fraction is still around 10\%.

\subsection{Stars with no spectroscopic solution}

\begin{figure}
    \centering
    \includegraphics[scale=0.57,trim={0.1cm 0.1cm 0.5cm 0.5cm},clip]{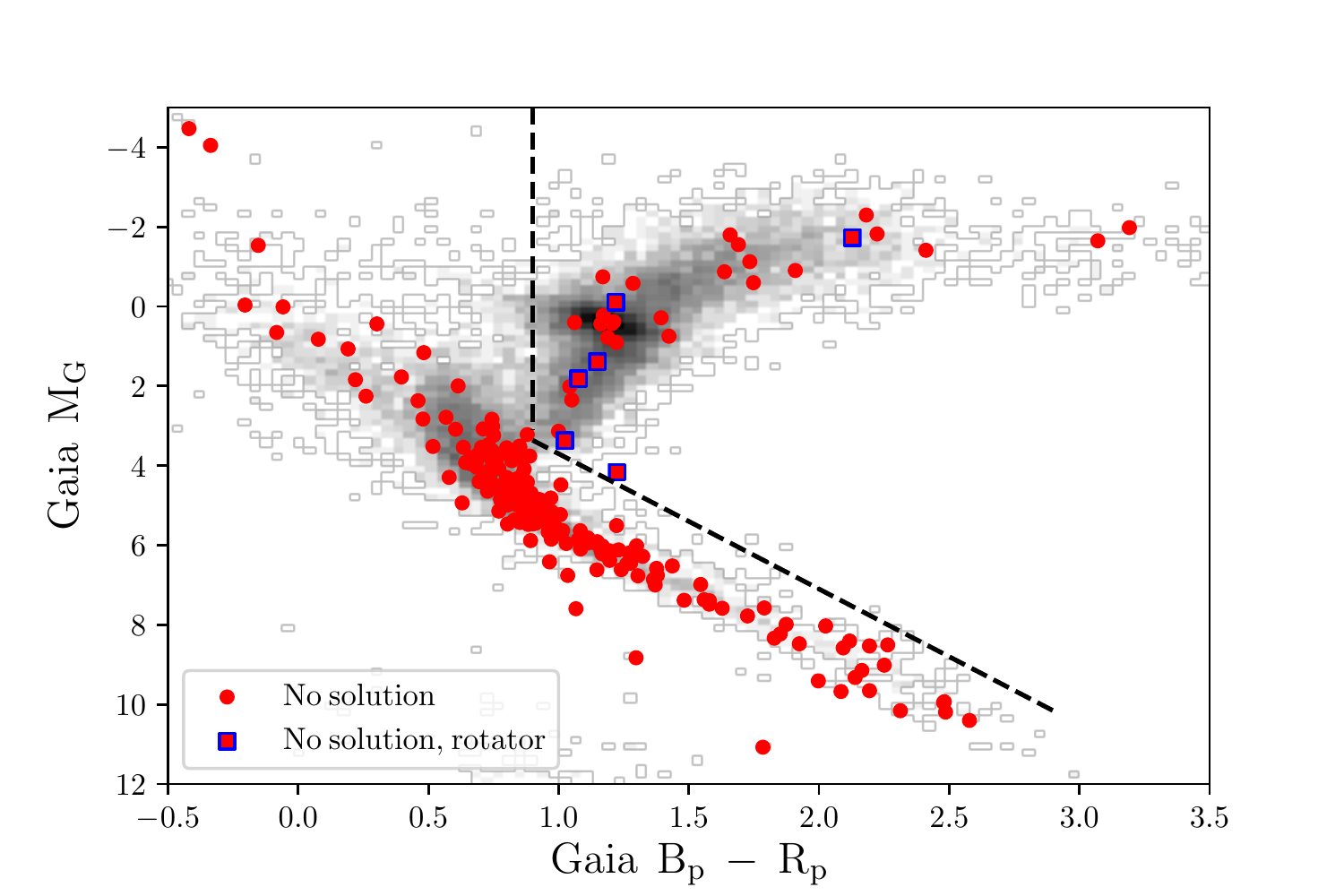}
    \caption{A \textit{Gaia} CMD of stars with no spectroscopic solution compared to the entire APOKASC-3 sample (grey). All points above and to the right of the dashed line we consider giants. Red stars are all targets with no solution. Red squares with a blue outline represent the stars which have other indicators of rapid rotation, such as rotational variability, but no \textit{v}sin\textit{i} measurement. While these giants are rapid rotators, they do not count toward our uniform assessment of rapid rotation.}
    \lFig{none}
\end{figure}

ASPCAP was unable to obtain solutions for some targets. Rotation could contribute to the poor quality for the same reasons outlined above, noisy spectra from activity, or combined spectral features from two stars in a binary.

The lack of spectroscopic information about these stars makes them the most difficult to study but points to the usefulness of the APOKASC-3 catalog. We still have the light curves of these stars to search for rotation signatures. However, we are limited by which of these targets show up in other surveys because we need external information in order to classify each star first as a red giant and then as a rapid rotator.  

There are 232 stars in APOKASC-3 without a spectroscopic solution. Table 3 lists the properties of these stars included in the table released with this paper. Cross matching with \textit{Gaia} to photometrically separate the red giants from main sequence stars, we end up with 32 giants. Their locations on a \textit{Gaia} CMD is shown in \Fig{none}. These stars do not appear to be photometric outliers, lying well within the red giant branch. Since these stars lack spectroscopic solutions, thus lack a best-fit template spectrum, we cannot measure their \textit{v}sin\textit{i}. However, \textit{Gaia} reports a $v_\mathrm{Broad}$ measurement for 18 targets, none of which are rapid rotators at the 20 km s$^{-1}$ threshold. 6 targets do show signs of rotational variability, so we do not think that giants in this category are devoid of rapid rotators. However, a uniform assessment of rapid rotation for this sample is difficult; these giants require significant follow-up. 

These stars are difficult to characterize because they lack APOGEE stellar parameters and radial velocity data. There are also very few of them. As such, we do not report binary fractions for giants in this category. While these stars may still contribute significantly to the total population of rapid rotators, a proper characterization, including an assessment of binarity, is needed.

\subsection{Rapid rotation in the complete set of APOKASC-3 giants}
\lSect{Apo3_rot}
\begin{figure}
    \centering
    \includegraphics[scale=0.57,trim={0.1cm 0.1cm 0.5cm 0.5cm},clip]{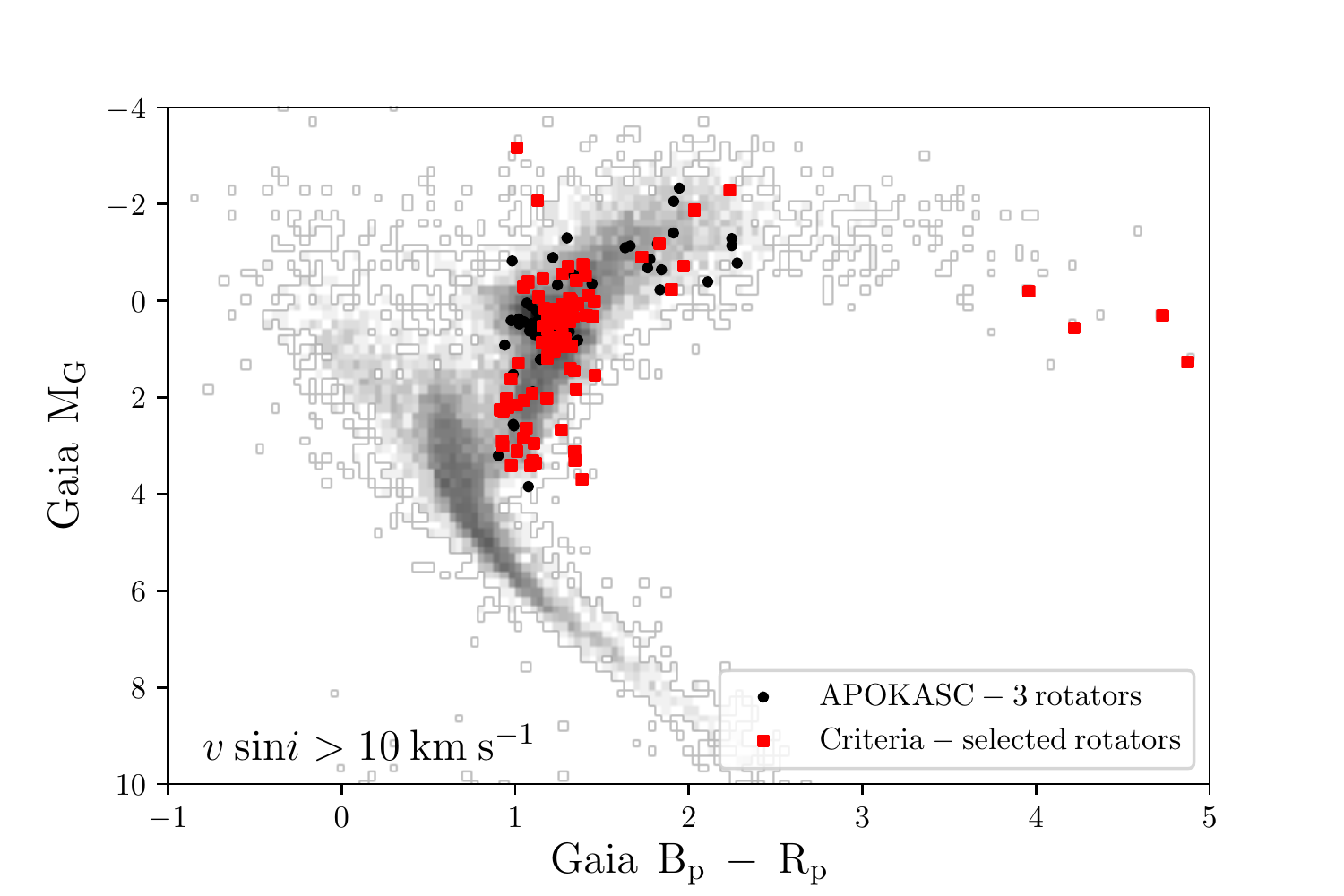}
    \includegraphics[scale=0.57,trim={0.1cm 0.1cm 0.5cm 0.5cm},clip]{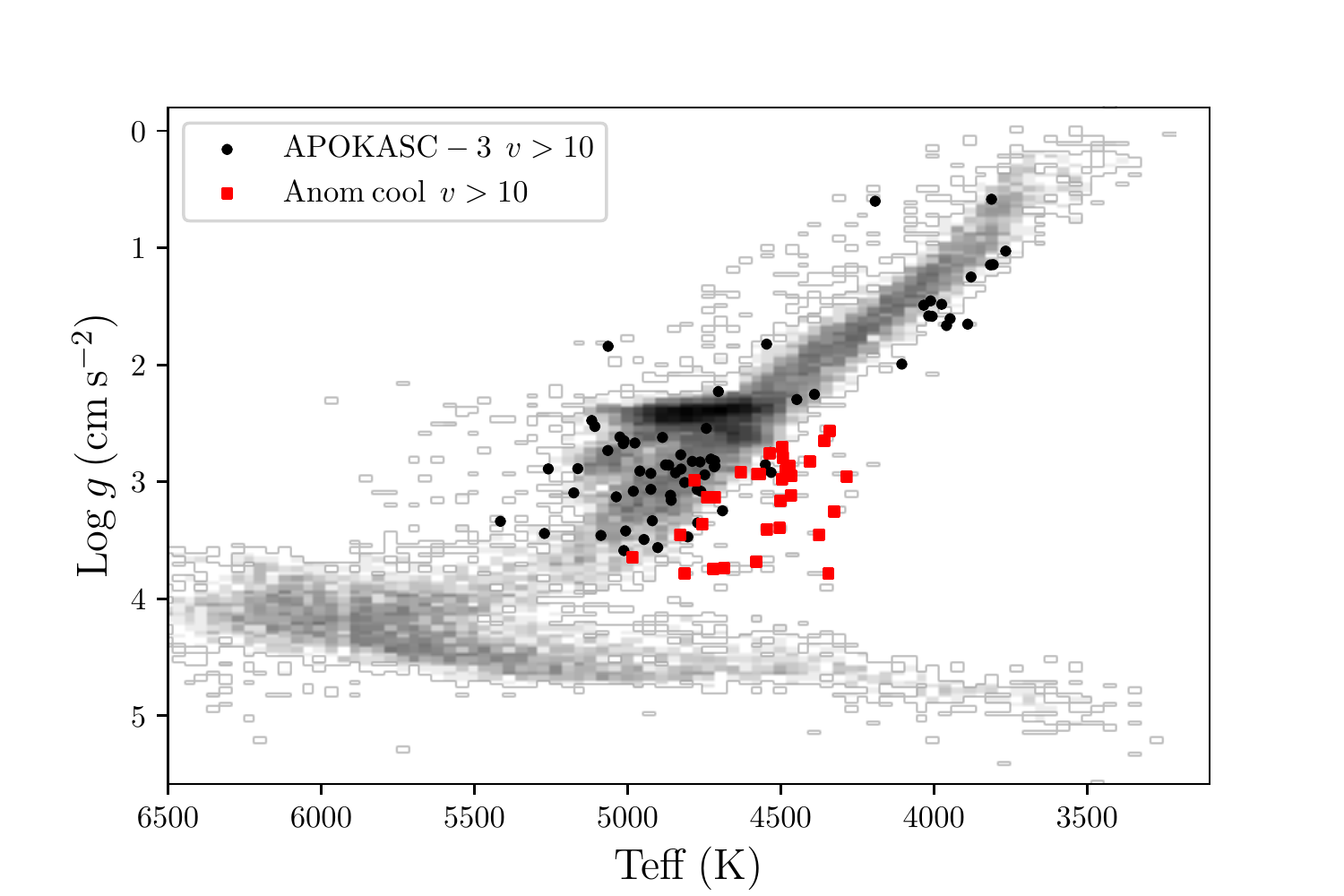}
    \caption{Top: A \textit{Gaia} CMD of the complete set of APOKASC-3 rotators with \textit{v}sin\textit{i} > 10 km s$^{-1}$(black) compared to stars with no spectroscopic solution compared to those extracted by our spectroscopic criteria (red). Bottom: a comparison of the spectroscopic stellar parameters for the complete set of APOKASC-3 rotators (black) with \textit{v}sin\textit{i} > 10 km s$^{-1}$ with those picked up in the anomalously cool criterion (red)}
    \lFig{Ap}
\end{figure}

\begin{figure}
    \centering
    \includegraphics[scale=0.57,trim={0.1cm 0.1cm 0.5cm 0.5cm},clip]{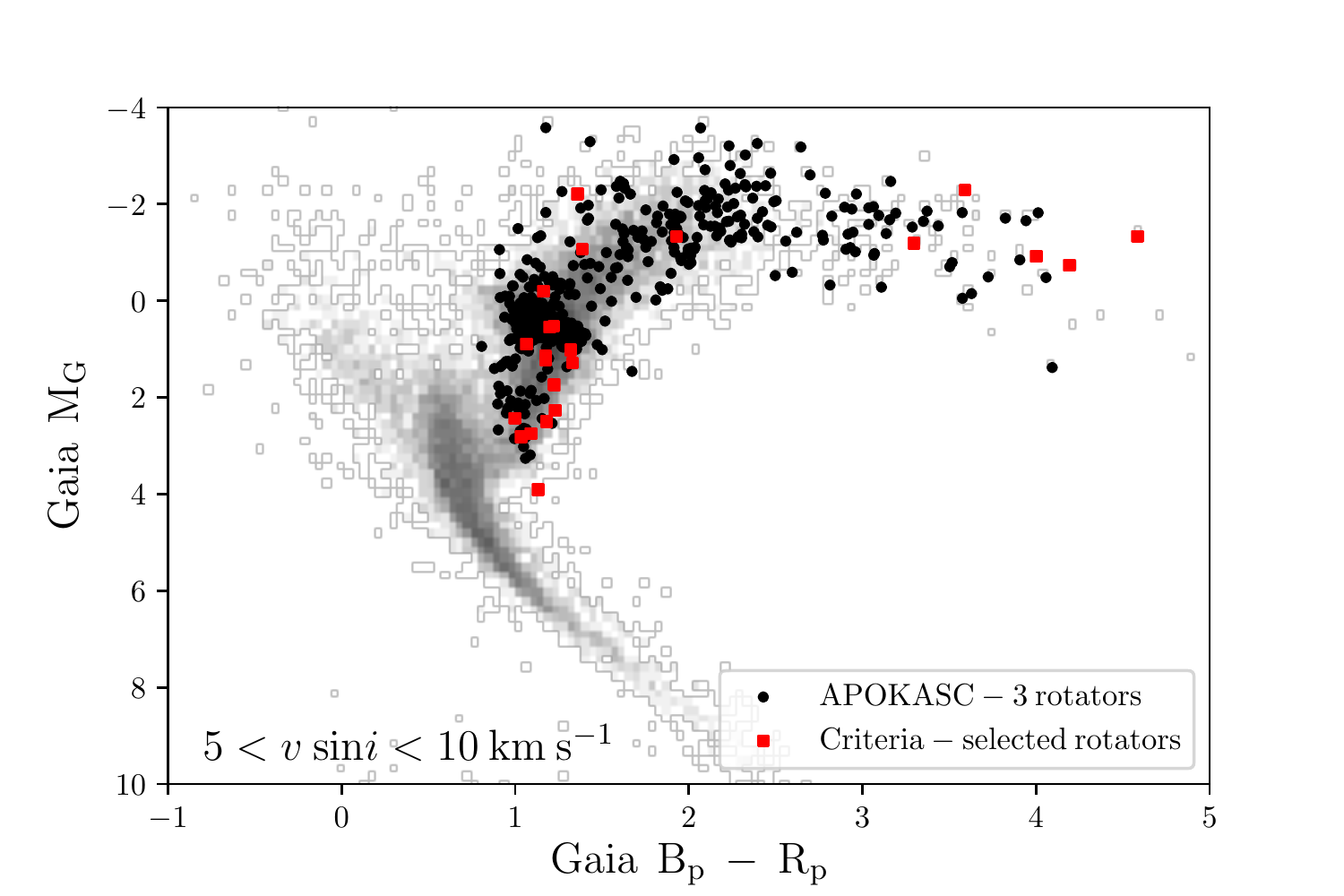}
    \includegraphics[scale=0.57,trim={0.1cm 0.1cm 0.5cm 0.5cm},clip]{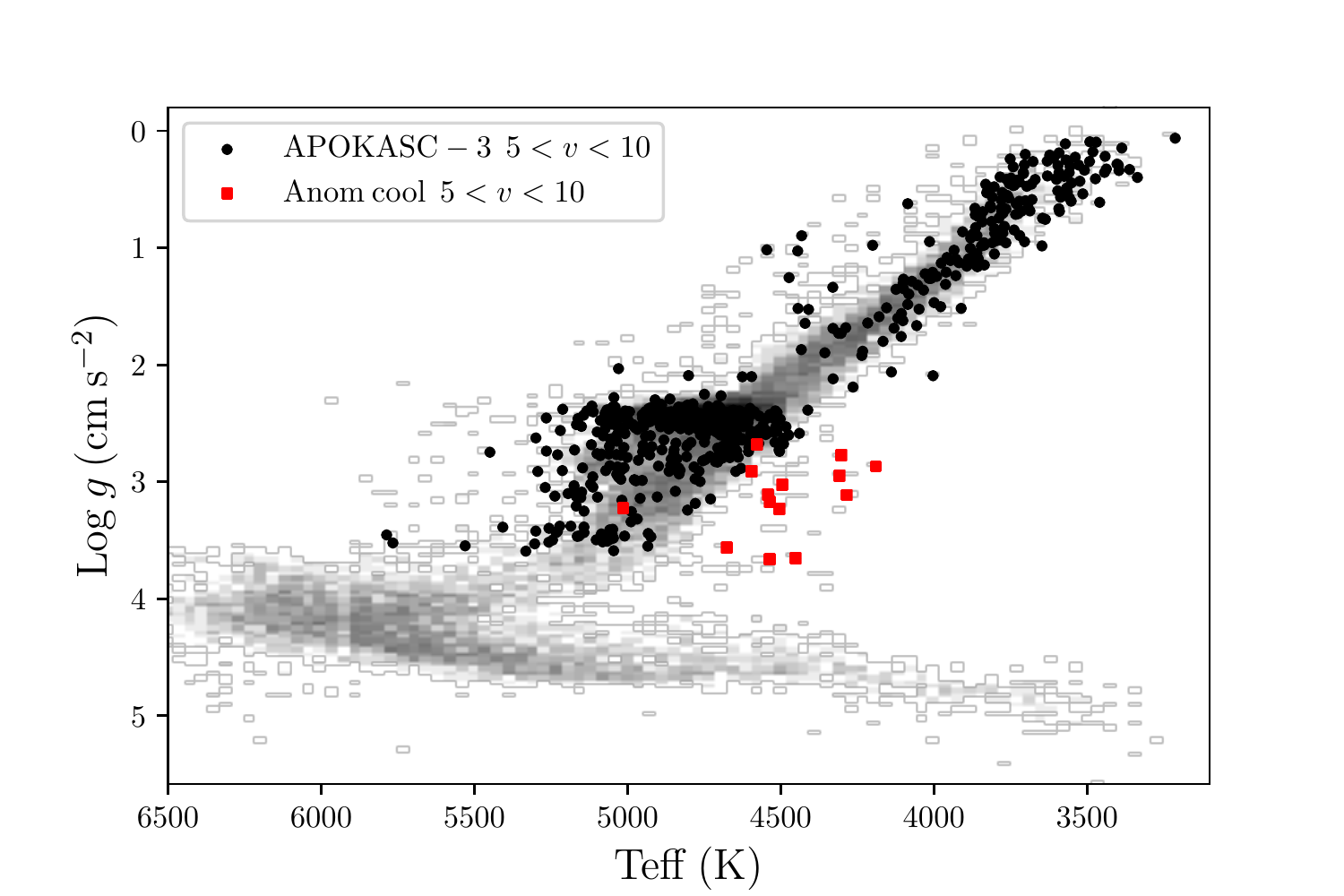}
    \caption{Top: A \textit{Gaia} CMD of the complete set of APOKASC-3 rotators with 5 < \textit{v}sin\textit{i} < 10 km s$^{-1}$(black) compared to stars with no spectroscopic solution compared to those extracted by our spectroscopic criteria (red). Bottom: a comparison of the spectroscopic stellar parameters for the complete set of APOKASC-3 rotators (black) with 5 < \textit{v}sin\textit{i} < 10 km s$^{-1}$ with those picked up in the anomalously cool criterion (red)}
    \lFig{Api}
\end{figure}

\begin{table*}
    \centering
    \caption{The counts of rapid rotators at the 5 and 10 km s$^{-1}$ threshold and their corresponding binary fractions broken down by evolutionary state. We lump together giants on the lower RGB where we could not separate the true first-ascent giants from the RC stars.}
    \begin{tabular}{lcccc}
        {} & Upper RGB & Lower RGB & RC & RGB/RC\\
        \hline
        \textit{v}sin\textit{i} > 10 km s$^{-1}$ & 28 & 41 & 34 & 48 \\
        RUWE f$_\mathrm{bin}$ & 18$\pm$8\% (5/28) & 24$\pm$8\% (10/41) & 3$\pm$3\% (1/34) & 17$\pm$6\% (8/48)\\
        \textit{Gaia} RV f$_\mathrm{bin}$ & 48$\pm$14\% (11/23)& 69$\pm$15\% (22/32) & 14$\pm$7\% (4/28) & 86$\pm$17\% (24/28)\\
        APOGEE f$_\mathrm{bin}$ & 85$\pm$26\% (11/13) & 71$\pm$16\% (20/28) & 73$\pm$26\% (8/11) & 62$\pm$15\%(18/29)\\
        APOGEE f$_\mathrm{bin, close}$ & 69$\pm$23\% (9/13) & 64$\pm$15\% (18/28) & 9$\pm$9\% (1/11) & 59$\pm$14\%(17/29)\\
        \hline
        5 < \textit{v}sin\textit{i} < 10 km s$^{-1}$ & 207 & 71 & 304 & 13 \\
        RUWE f$_\mathrm{bin}$ & 17$\pm$3\% (36/207) & 23$\pm$6\% (16/70) & 13$\pm$2\% (40/304) & 23$\pm$13\% (3/13)\\
        \textit{Gaia} RV f$_\mathrm{bin}$ & 9$\pm$2\% (16/172) & 42$\pm$9\% (23/55) & 14$\pm$2\% (35/247) & 23$\pm$13\% (3/13)\\
        APOGEE f$_\mathrm{bin}$ & 74$\pm$9\% (69/93) & 66$\pm$13\% (25/38) & 42$\pm$6\% (59/139) & 30$\pm$17\% (3/10)\\
        APOGEE f$_\mathrm{bin, close}$ & 58$\pm$8\% (54/93) & 50$\pm$11\% (19/38) & 11$\pm$3\% (15/139) & 10$\pm$10\% (1/10)\\
        \hline
    \end{tabular}
\end{table*}

To assess the yields and completeness of the criteria, we must know what the true underlying population of rapid rotators is in the APOKASC-3 sample. Measurements of \textit{v}sin\textit{i} can automatically confirm whether a star is rapidly rotating. Using the same methods as in \citet{Tay15}, \citet{Dix20}, and \citet{Dah22}, we determine \textit{v}sin\textit{i} by cross-correlating the combined DR16 spectra for 15 220 giants with their respective best-fit template and an artificially broadened version of that template. This only works for stars with a best-fit template spectrum. We cannot estimate \textit{v}sin\textit{i} for stars with no spectroscopic solution. While \citet{Tay15} adopts 10 km s$^{-1}$ as their threshold for rapid rotation, they find that stars with 5 < \textit{v}sin\textit{i} < 10 km s$^{-1}$ are rotationally enhanced, likely from binary interaction. The same is seen more broadly in the field in \citet{Dah22}. As such, we adopt 5 km s$^{-1}$ as the threshold for rapid rotation but maintain a distinction between stars with v > 10 km s$^{-1}$ and those with 5 < \textit{v}sin\textit{i} < 10 km s$^{-1}$. We employ the same photometric cuts as for stars identified by our spectroscopic selection criteria: \textit{Gaia} B$_\mathrm{P}$ - R$_\mathrm{P}$ color > 0.9 mag and absolute \textit{Gaia} G-band magnitude > 0.3 + 3.4*(dereddened B$_\mathrm{P}$ - R$_\mathrm{P}$).

Because the APOKASC-3 giants span the giant branch and the clump, a constant \textit{v}sin\textit{i} threshold for rapid rotation corresponds to vastly different rotation periods depending on the star's radius. As such, we break down the rapid rotators by their evolutionary state. Upper giant branch stars are defined as giants with log \textit{g} < 2.3 cm s$^{-2}$ or, for giants without complete solutions, \textit{Gaia} B$_\mathrm{P}$ - R$_\mathrm{P}$ color > 1.5 mag. For giants on the lower giant branch we separate the first-ascent giants from the clump using seismically-determined evolutionary states. For stars lacking seismology we adopt the spectroscopic evolutionary state determined in DR16. Table 5 shows the breakdown of rapid rotators by evolutionary state as well as their respective binary fractions from \textit{Gaia} and APOGEE data.

We demonstrated that our spectroscopic selection criteria have a high purity but we still need to assess completeness. Our method selects for stars with rotation, or activity, substantial enough to produce anomalies in the spectroscopic solution. Furthermore we focus on the lower RGB, where these anomalies are the largest. Slower rotators, or those in other regimes, will not be detected with our technique. 

In APOKASC-3, we detect 151 stars with \textit{v}sin\textit{i} > 10 km s$^{-1}$, and our spectroscopic criteria combined capture 85 (56\%) of them. However, on the lower RGB, where our anomalously cool and partial solution criteria are most sensitive, we capture 33 (80\%) of the 41 confirmed lower RGB rapid rotators. Still, we see rapid rotators outside the evolutionary states where we are sensitive. Of the 67 stars we do not recover, 19 are confirmed on the upper RGB, 8 are on the lower RGB, and the remaining 40 are either in the clump or were a combination of first-ascent and clump giants. \Fig{Ap} shows that our spectroscopic selection criteria are efficient in recapturing the rapid rotators with v > 10 km s$^{-1}$. 

We observe 595 giants with 5 < \textit{v}sin\textit{i} < 10 km s$^{-1}$. Our results are consistent with \citet{Tay15}, who found that stars with \textit{v}sin\textit{i} in this range significantly outnumbered the traditional rapid rotators. Even when binaries are targeted first and then rapid rotation is assessed, the trend holds \citep{Dah22}. The vast majority of these targets are found in the clump or the upper RGB, seen most clearly in the bottom panel of \Fig{Api}. As such, we reach the same conclusions as \citet{Tay15}, namely that these stars are interacting, and likely merging, on the upper RGB and retaining enhanced rotation in the clump as a result. This is supported by the low binary fraction of these intermediate rotators in the clump. Because the majority of these targets do not live on the lower RGB, where our criteria are most sensitive, we only recover 26 of these rotators.

Two things are apparent. The first is that our spectroscopic selection criteria preferentially pick out the most rapidly rotating giants, suggesting that higher rotation speeds are needed to induce spectroscopic anomalies. The second is that population of rotationally enhanced giants is far larger than we expected. Ultimately, 4.9\% of the APOKASC-3 catalog is rapidly rotating at the 5 km s$^{-1}$ threshold, implying a previously unidentified population of giants in the \textit{Kepler} field which have experienced binary interaction.

As an aside, we call attention to the population of rotators to the left of the giant branch in \Fig{Ap} and \Fig{Api}. These stars occupy a well-defined region of log \textit{g} - $T_\mathrm{eff}$ space, namely 2.3 < log \textit{g} < 3.85 cm S$^{-2}$and 5300 - 6000 K. There are only a few stars in this region because we impose a photometric cut at \textit{Gaia} B$_\mathrm{P}$ - R$_\mathrm{P}$ = 0.9. Without this cut, this region is much more populated with rapid rotators. These stars are likely crossing the Hertzsprung gap. We suggest the above spectroscopic criteria for identifying these targets in APOGEE.

\subsection{Binary Origins of Rotation in APOKASC-3}

The binary fraction of giants should change as a function of evolutionary state \citep{Bad18,Pri20}. As giants expand up the RGB, systems which were tidally synchronized on the lower RGB should merge on the upper RGB. This then leads to a lower binary fraction in the clump. This is mostly reflected in the binary fractions in the entire set of APOKASC-3 rotators; the binary fraction are comparable between the lower and upper RGB but overall lower in the clump. 

The giants with partial solutions reflect this distribution of binaries, which makes sense since these giants span the RGB and clump. The anomalously cool giants have higher binary and close binary fractions, suggesting that most of these stars are in close binaries on the lower RGB.  

\section{Rapidly rotating red giants in APOGEE DR16}
\lSect{dr16}
\subsection{Assembling the catalog}
\begin{figure}
    \centering
    \includegraphics[scale=0.55,trim={0.0cm 0.1cm 0.5cm 0.5cm},clip]{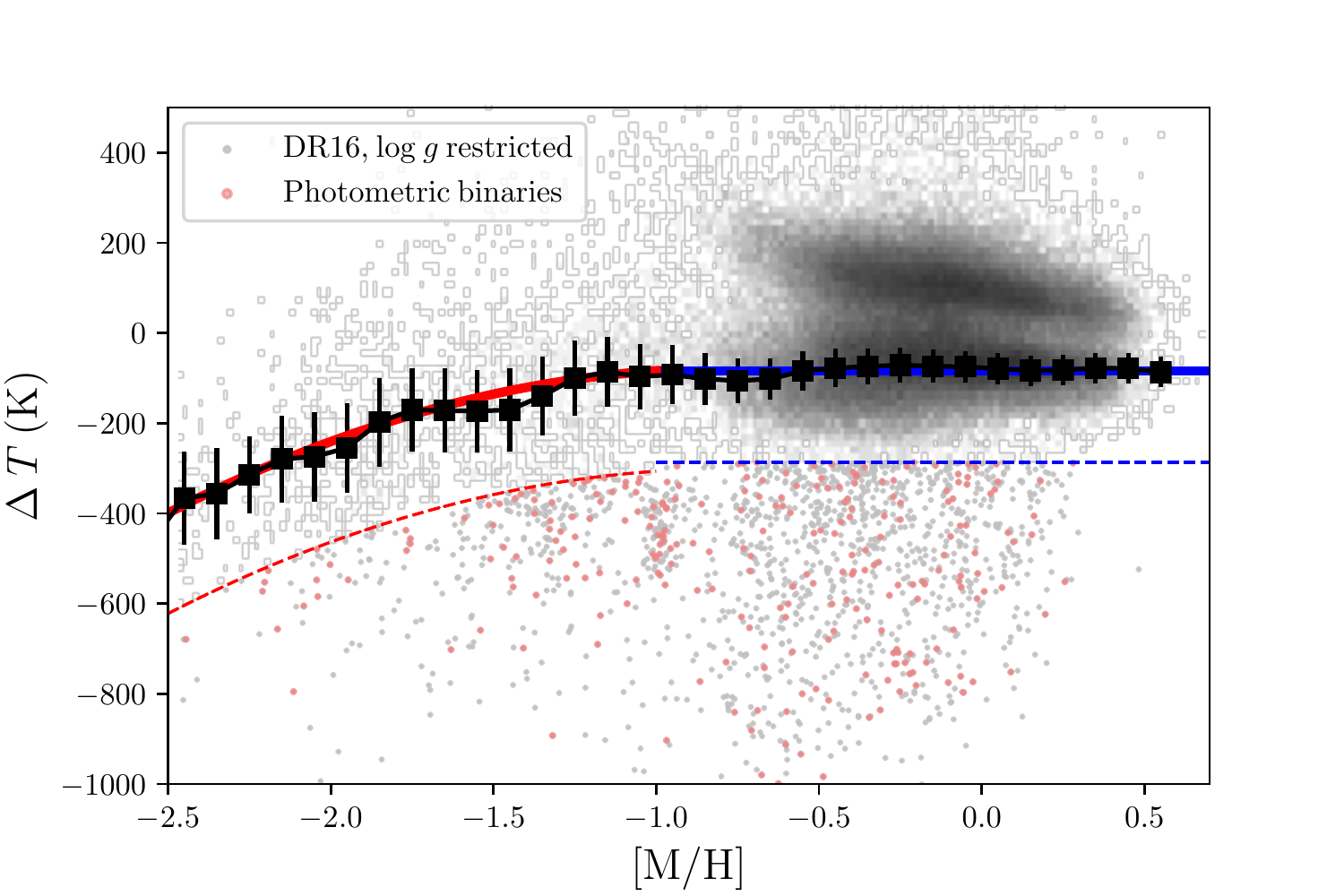}
    \caption{Calculated $\Delta T$ for all DR16 stars (grey) with 2.3 < log \textit{g} < 3.85 versus metallicity. Photometric binaries are highlighted in pink. They do not contribute to our final counts. The solid lines represent the fits to the binned data (black squares), with blue corresponding to the high metallicity population and red corresponding to the low metallicity population. Dashed lines represent the 3$\sigma$ threshold, below which stars are deemed anomalously cool. We limit the y-axis to focus on the main group of anomalously cool stars, though some targets have much lower $\Delta T$.}
    \lFig{dr16_cool_abun}
\end{figure}

\begin{figure}
    \centering
    \includegraphics[scale=0.6,trim={0.5cm 0.1cm 0.5cm 0.5cm},clip]{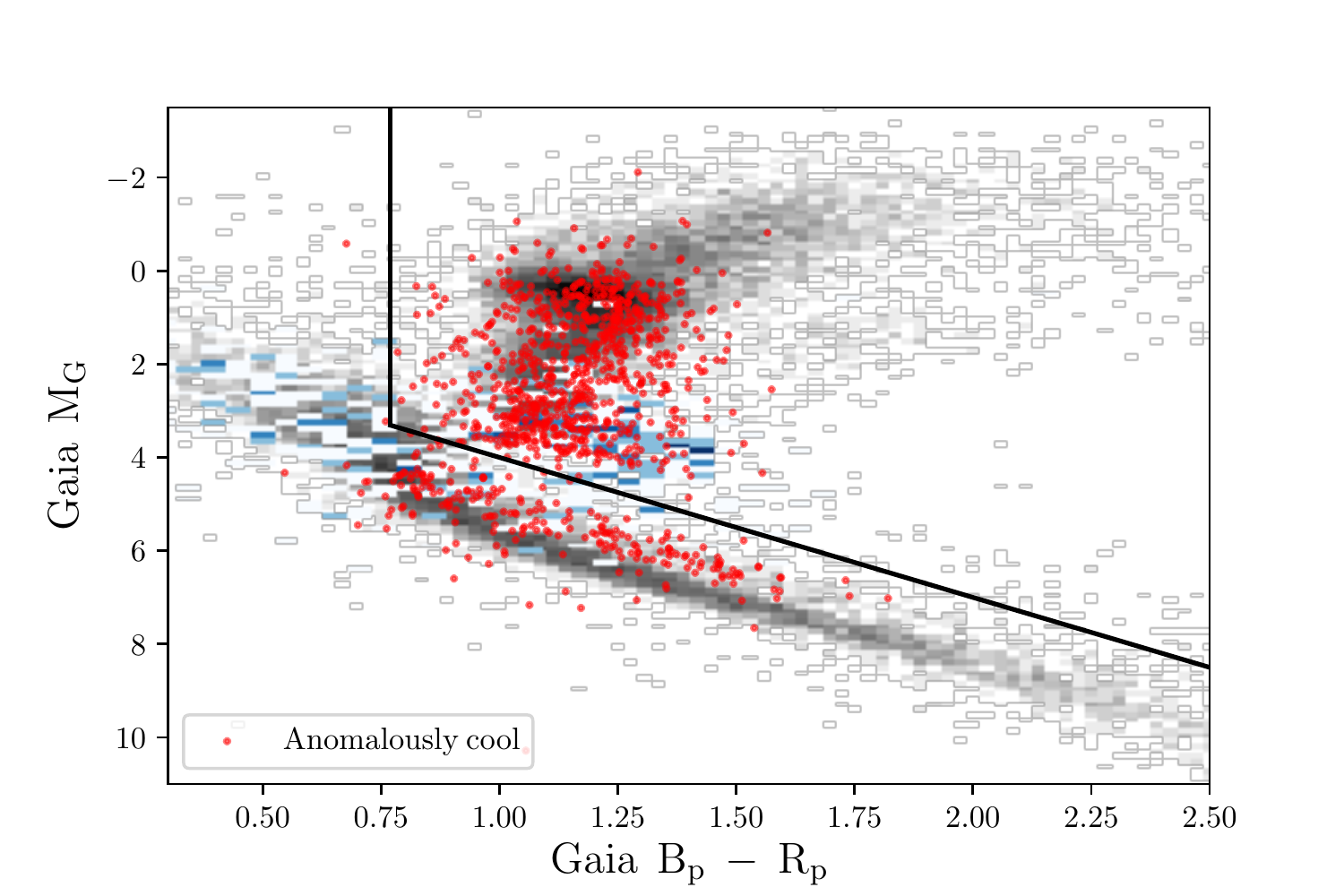}
    \caption{A \textit{Gaia} CMD of a subset of DR16 stars (grey) with the anomalously cool stars highlighted in red. The blue distribution represents the location of a sample of RS CVn-like variables, which preferentially occupy the subgiant and lower giant branch. The black line represents color-magntiude cuts we make to remove dwarfs from our sample.}
    \lFig{dr16_cool_CMD}
\end{figure}

\begin{figure}
    \centering
    \includegraphics[scale=0.6,trim={0.5cm 0.1cm 0.5cm 0.5cm},clip]{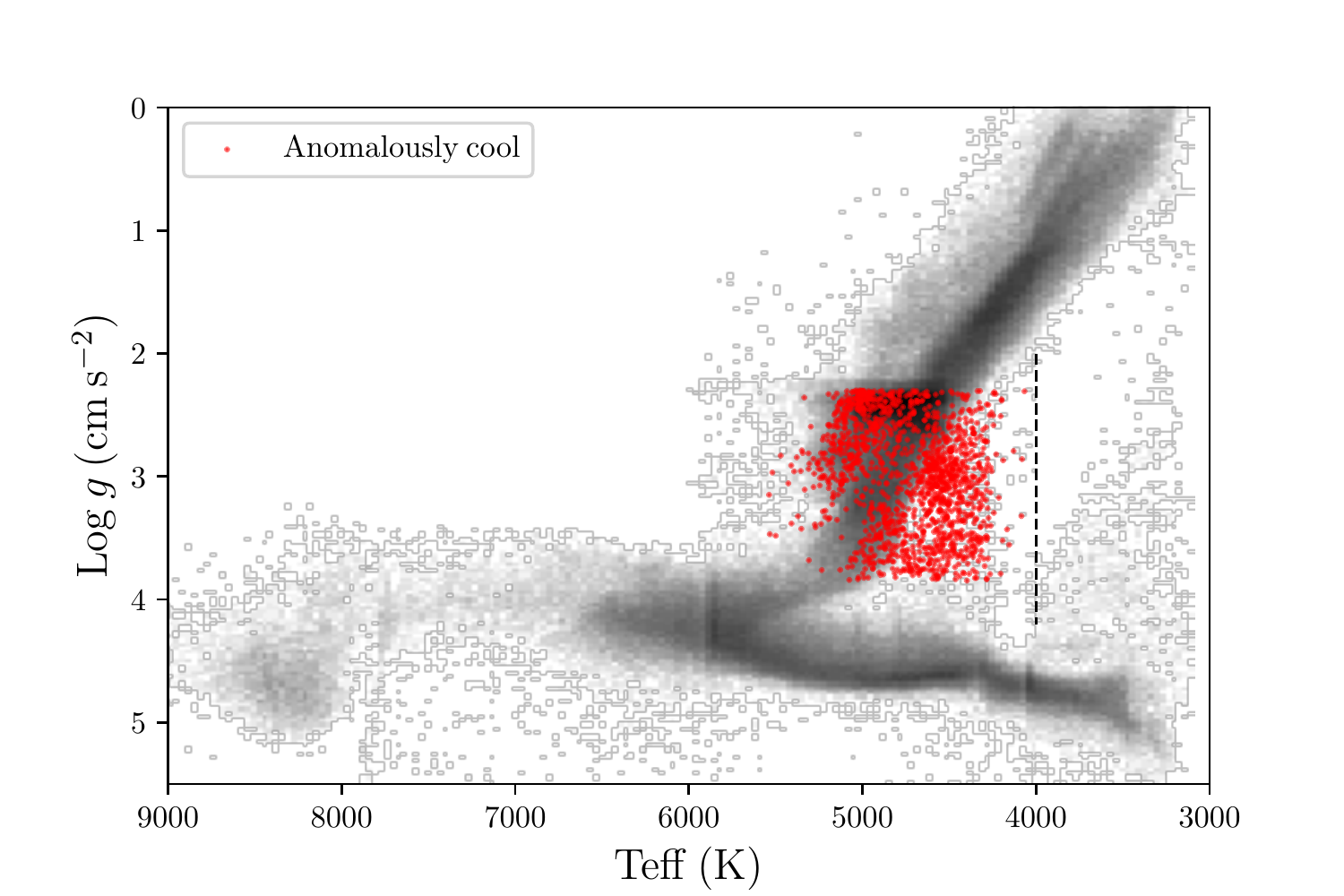}
    \caption{Calibrated log \textit{g} versus $T_\mathrm{eff}$ for the entire DR16 sample (grey) with the anomalously cool stars highlighted in red. YSOs are visible above the main sequence with temperatures below 4000 K, our adopted cutoff (dashed line).}
    \lFig{dr16_cool}
\end{figure}

\begin{figure}
    \centering
    \includegraphics[scale=0.6,trim={0.5cm 0.1cm 0.5cm 0.5cm},clip]{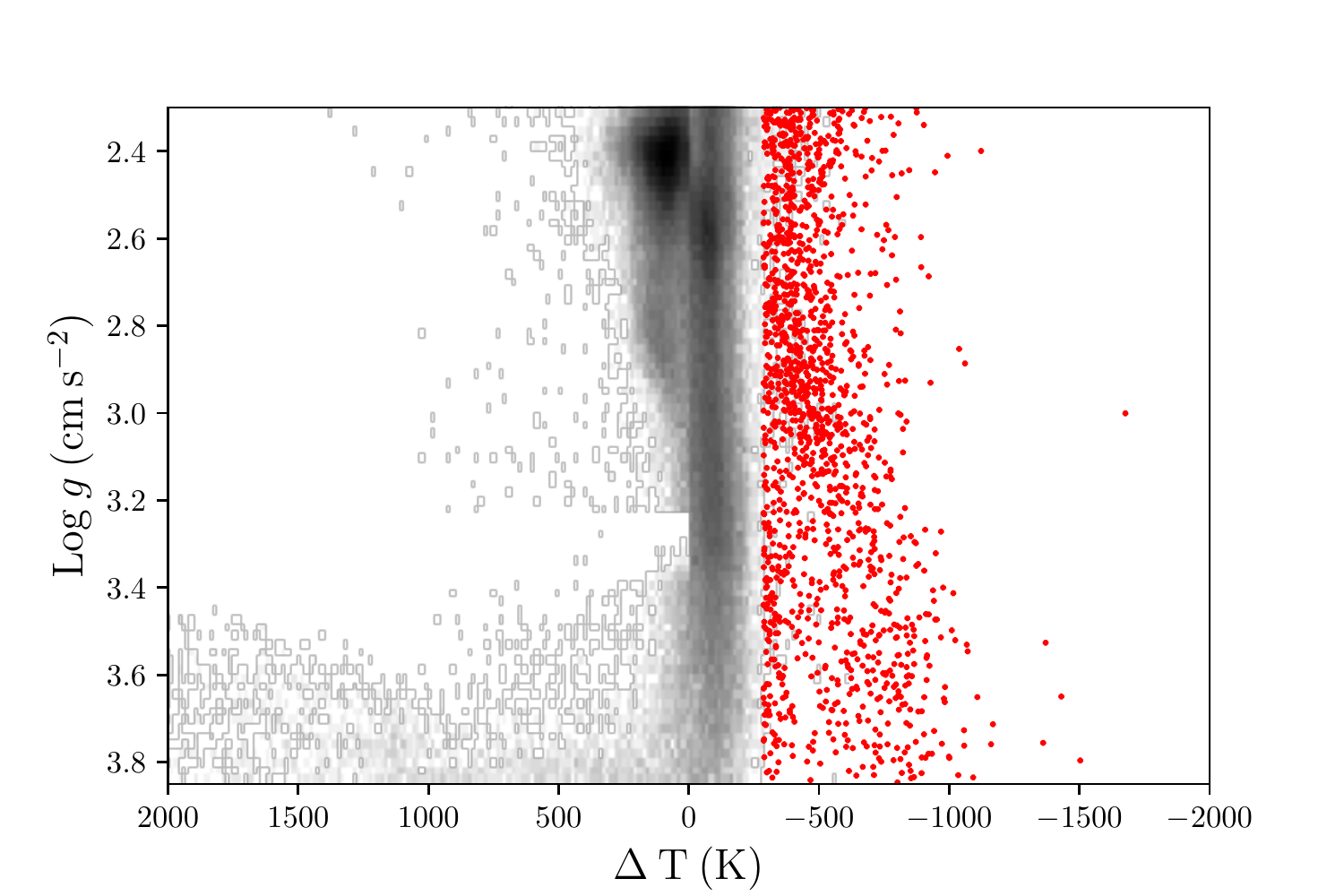}
    \caption{Log \textit{g} versus $\Delta T$ for log \textit{g}-restricted set DR16 stars (grey) and the anomalously cool stars (red). The anomalously cool stars are distinct from the rest of DR16.}
    \lFig{dr16_cool_dt}
\end{figure}

\begin{figure}
    \centering
    \includegraphics[scale=0.55,trim={0.0cm 0.1cm 0.5cm 0.5cm},clip]{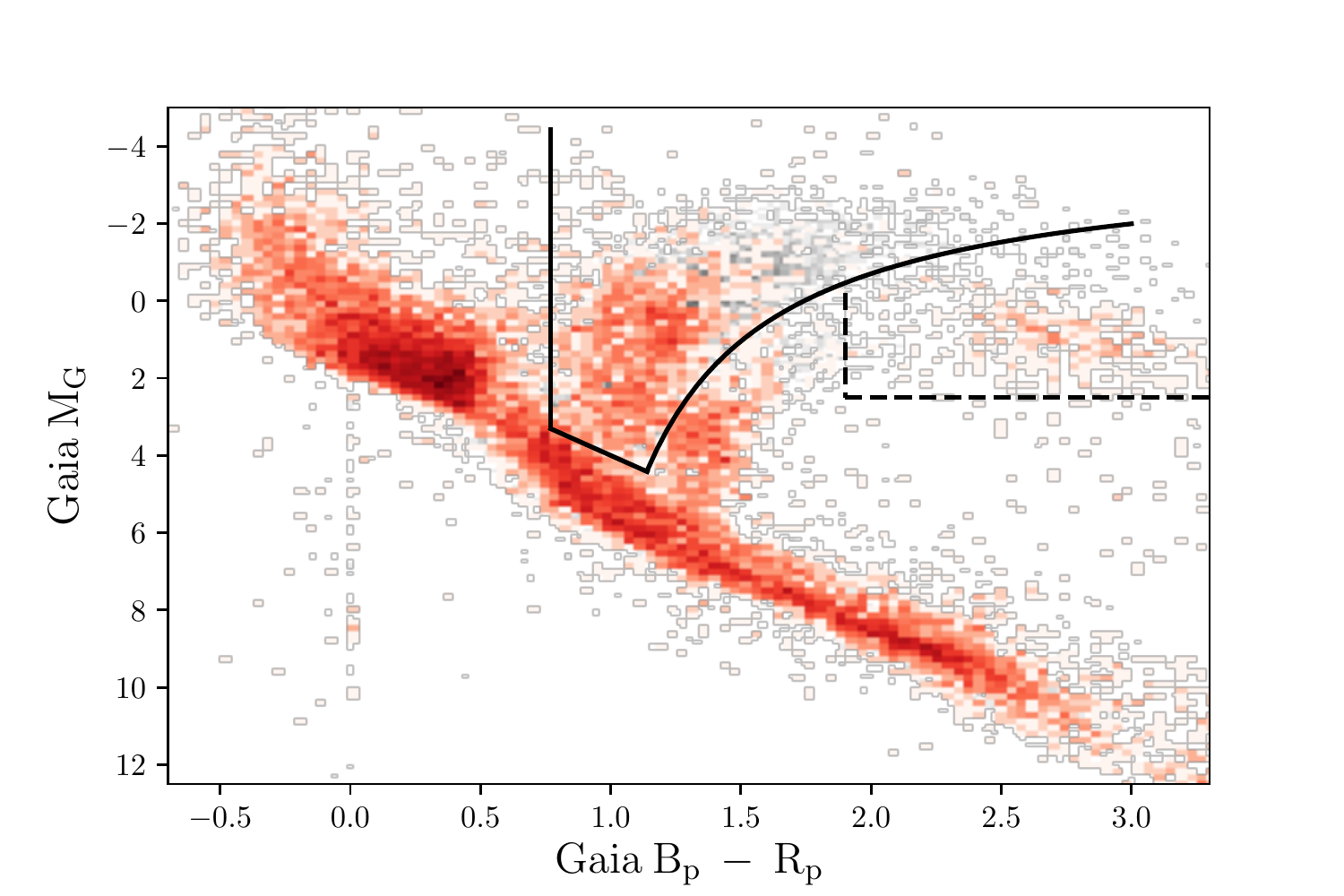}
    \caption{A \textit{Gaia} DR2 CMD for a subset of the APOGEE DR16 sample (grey) and stars with partial spectroscopic solutions (red). Black lines denote the photometric cuts we impose to isolate the giants.}
    \lFig{dr16_part}
\end{figure}

\begin{table}
  \caption{Column names for the table containing our APOGEE DR16 anomalously cool rapid rotator candidate sample}
  \lTab{dr16_cool}
  \centering
  \begin{tabular}{lr}
  \\
  \hline
  APOGEE ID & APOGEE DR16\\
  \textit{Gaia} ID & \textit{Gaia} DR3\\
  log \textit{g} (cm s$^{-2}$) & APOGEE DR16\\ 
  $T_\mathrm{eff}$ (K) & APOGEE DR16\\
  {[M/H]} & APOGEE DR16\\
  $\Delta T$ (K) & this work\\
  m$_\mathrm{G}$ & \textit{Gaia} DR2 \\
  m$_{\mathrm{B}_\mathrm{P}}$ & \textit{Gaia} DR2\\
  m$_{\mathrm{R}_\mathrm{P}}$ & \textit{Gaia} DR2\\
  $A_V$ & this work\\ 
  distance (pc) & \textit{Gaia} DR3\\
  \textit{v}sin\textit{i} (km s$^{-1}$) & this work\\
  $v_\mathrm{Broad}$ (km s$^{-1}$) & \textit{Gaia} DR2\\
  RUWE & \textit{Gaia} DR3\\
  \textit{Gaia} RV variable & this work\\
  APOGEE RV variable & this work\\
  \hline
  \end{tabular}
\end{table}

\begin{table}
  \caption{Column names for the table containing our APOGEE DR16 rapid rotator candidate sample of giants with partial solutions}
  \lTab{dr16_other}
  \centering
  \begin{tabular}{lr}
  \\
  \hline
  APOGEE ID & APOGEE DR16\\
  \textit{Gaia} ID & \textit{Gaia} DR3\\
  m$_\mathrm{G}$ & \textit{Gaia} DR2 \\
  m$_{\mathrm{B}_\mathrm{P}}$ & \textit{Gaia} DR2\\
  m$_{\mathrm{R}_\mathrm{P}}$ & \textit{Gaia} DR2\\
  $A_V$ & this work\\ 
  distance (pc) & \textit{Gaia} DR3\\
  \textit{v}sin\textit{i} (km s$^{-1}$) & this work\\
  $v_\mathrm{Broad}$ (km s$^{-1}$) & \textit{Gaia} DR2\\
  RUWE & \textit{Gaia} DR3\\
  \textit{Gaia} RV variable & this work\\
  APOGEE RV variable & this work\\
  Flags & APOGEE DR16\\
  \hline
  \end{tabular}
\end{table}

Our total candidate list contains 6933 targets, including 1484 anomalously cool giants, 1733 giants with partial solutions (plus 340 luminous giants with partial solutions), and 3376 giants with no solution. We discuss each category below and follow up with \textit{v}sin\textit{i} confirmation in \Sect{dr16_conf}.

To apply the anomalously cool criterion to the entire APOGEE DR16 sample, we must impose an additional cut that was not necessary in the APOKASC-3 sample due to it being an older population. Some young stellar objects (YSOs) fall in the same log \textit{g} range as the lower giant branch but have lower effective temperatures. The YSO population is very cleanly separated from the red giants, as shown in \Fig{dr16_cool}, and we can remove them with a simple $T_\mathrm{eff}$ cut at 4,000 K and below.

We must also make corrections to account for the different metallicity ranges covered by DR16 and APOKASC-3. The evolutionary state classifier, which provides the basis for finding anomalously cool stars, was calibrated to the \textit{Kepler} field, where metallicity extends down to [M/H] of -2.46, with the vast majority of stars having metallicity above -0.5. There is therefore no guarantee that the \textit{Kepler} relationships will extrapolate to the low metallicity regime. 

We see that they do not. \Fig{dr16_cool_abun} shows the distribution of all stars in DR16 with 2.3 < log \textit{g} < 3.85 cm s$^{-2}$ in $\Delta T$ - metallicity space. By definition, all RGB stars have $\Delta T < 0$. Down to [M/H] of about -1 RGB stars tend to lie in an even distribution centered on $\Delta T = -83.6$K. At lower metallicity, the RGB stars trend downward in $\Delta T$.

To identify anomalously cool stars at low metallicity, we first bin the stars with metallicity between -2.4 and -1 in increments of 0.1 dex and iteratively reject stars which are more than 3$\sigma$ from the median $\Delta T$ in each bin. Here $\sigma$ represents the absolute median deviation. Red clump stars have $\Delta T$ values above zero, and could in principle bias the results. This contamination is mitigated by the usage of median statistics and by stellar population effects. The majority of core He burning stars in this metallicity domain are hotter than 6000 K and therefore outside of the domain considered here. We then fit a second-order polynomial to the medians of each bin, the solid red line in \Fig{dr16_cool_abun}. To align the median lines of the red giants above and below [M/H] = -1, we apply an offset of +3.24 K to the polynomial. Finally, we define a boundary at 3$\sigma$ to isolate the anomalously cool stars. The dashed lines in \Fig{dr16_cool_abun} represent the 3$\sigma$ boundary, below which all stars are flagged as anomalously cool. Above [M/H] of -1, the line is fixed at -287K, and below it is metallicity dependent. We discuss the metallicity dependence of our yields in Section 4.2. 

We then remove 310 candidates that are not giants by excluding all targets with absolute \textit{Gaia} G-band magnitude M$_G$ > 1 + 3*\textit{Gaia} B$_\mathrm{P}$ - R$_\mathrm{P}$ and \textit{Gaia} B$_\mathrm{P}$ - R$_\mathrm{P}$ > 0.77 mag. As in the APOKASC-3 sample, these stars lie along the photometric binary sequence, and we believe that they represent a failure mode where the combined light of two stars contributing significant flux yields inaccurate stellar parameters. In total, our anomalously cool sample consists of 1,484 giants. Like the APOKASC-3 sample, these stars are not actually redder than typical giants. A \textit{Gaia} CMD of the anomalously cool stars and a subset of the entire DR16 sample, \Fig{dr16_cool_CMD}, shows that most stars fall comfortably on the lower RGB. We note that this list includes 82 targets that we could not verify as giants because of either very large formal $A_V$ values or unreliable distances. Table 6 lists the column names of the table containing the properties of these giants.

Active field giants are preferentially found on the lower RGB. Many active giants are in synchronized binaries, and while synchronization timescales increase steeply with orbital period, evolutionary timescales decrease with luminosity. On the upper RGB, stars do not have enough time to synchronize with their companion before coming into contact. Observations of active giants confirm this. RS CVn type variables, a binary where the primary is an active giant or subgiant, live predominantly on the lower RGB with orbital periods mostly below the synchronization limit of 20 days (Fig 7 in \citet{Lei21}). Since we expect many of our targets to be active giants in binaries, RS CVn candidates, they should fall in the same CMD location as the RS CVns. We can see from the overlap of our targets with the blue distribution in \Fig{dr16_cool_CMD}, the CMD location of a sample of about 550 known RS CVns, that they do. 

In spectroscopic parameter space, our sample looks like the APOKASC-3 sample. \Fig{dr16_cool} shows that these stars, mostly, but not completely, stand out from typical red giants. In log \textit{g} - $\Delta T$ space, these stars are completely isolated from the rest of the sample, shown in \Fig{dr16_cool_dt}. Like the APOKASC-3 sample, the majority of stars have temperature offsets between -1000 and -287 K, though there are some which extend to higher offsets. The value of $\Delta T$ does not correlate with the rotation speed, so it remains unclear as to why some stars have such large $T_\mathrm{eff}$ offsets from a typical red giant.

Stars with partial spectroscopic solutions are more difficult to identify because they are not isolated to one evolutionary state. There are 31821 stars spread out across a \textit{Gaia} CMD (\Fig{dr16_part}). To isolate the stars along the RGB, we start with the stars with the best characterized photometry, excluding any target with an $A_\mathrm{V} >1$. We then impose 3 cuts: a cut at a \textit{Gaia} B$_\mathrm{P}$ - R$_\mathrm{P}$ of 0.77 to exclude hot main sequence stars, a diagonal cut at the turnoff, the same as in the anomalously cool sample, to exclude the main sequence, and a cut along the red edge of the RGB to exclude YSOs. In total, we identify 1733 rapidly rotating giant candidates. Table 7 lists the properties of these targets we include in the catalog released with the paper. Combining the targets identified by both spectroscopic criteria, our main catalog ultimately contains 3217 rapid rotator candidates. 

In the top-right section of \Fig{dr16_part}, there is a population of 340 luminous giants excluded by our photometric cuts. Our independent spectroscopic fitting method is not sensitive to these very red stars, so we cannot assess rotation. Furthermore, these giants require a more careful analysis due to more sources of variability, including Mira-like variables and AGB thermal pulses, which could contribute to their poorer spectral quality. We also see in \Fig{partial} and \Fig{Ap} that in the APOKASC-3 catalog these stars are mostly devoid of rapid rotators, which our criteria are most sensitive to. As such, we include the APOGEE IDs of these stars in a separate table but do not include them in our main catalog. 

There are 3376 stars with no spectroscopic solution, selected via the same photometric cuts as the stars with partial solutions. In APOKASC-3, we did not have a uniform, definitive means to assess the fraction of these stars which exhibit rapid rotation. Quality control is difficult for stars with no spectroscopic solution, so we do not include these in our main sample. However, we do include the APOGEE IDs of targets in a separate table, as these targets likely contain rapid rotators.

\subsection{Confirming rapid rotation}
\lSect{dr16_conf}

\begin{figure}
    \centering
    \includegraphics[scale=0.55,trim={0.0cm 0.1cm 0.5cm 0.5cm},clip]{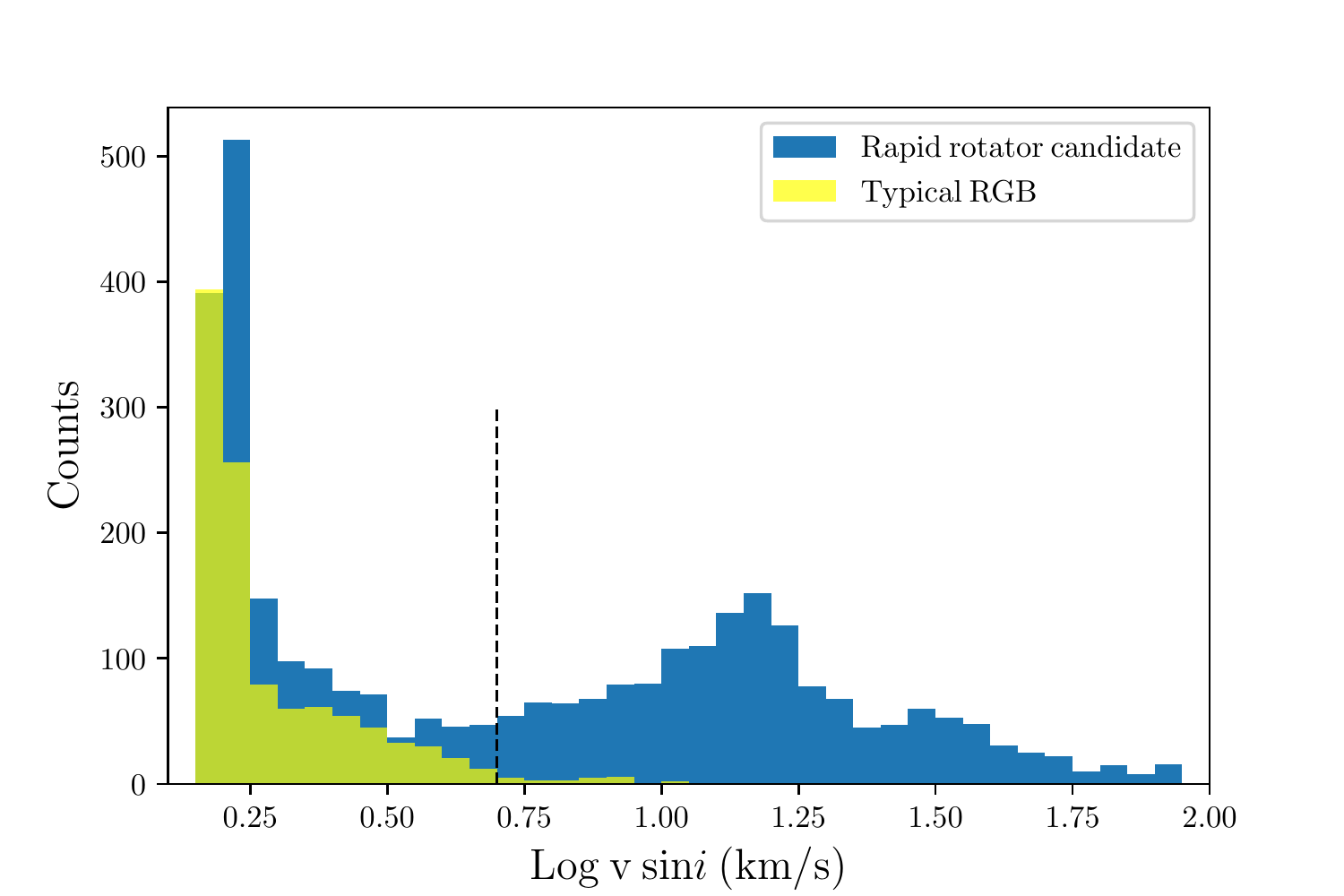}
    \caption{The distribution of \textit{v}sin\textit{i} measurements from our independent spectral fits for our rapid rotator candidates (blue) and a subset of typical red giants (yellow) in APOGEE. The vertical dashed line indicates the uptick in rotation starting at log \textit{v}sin\textit{i} of 0.7, corresponding to a distinct population of rapid rotators.}
    \lFig{lyra}
\end{figure}

\begin{figure}
    \centering
    \includegraphics[scale=0.56,trim={0.0cm 0.1cm 0.5cm 0.5cm},clip]{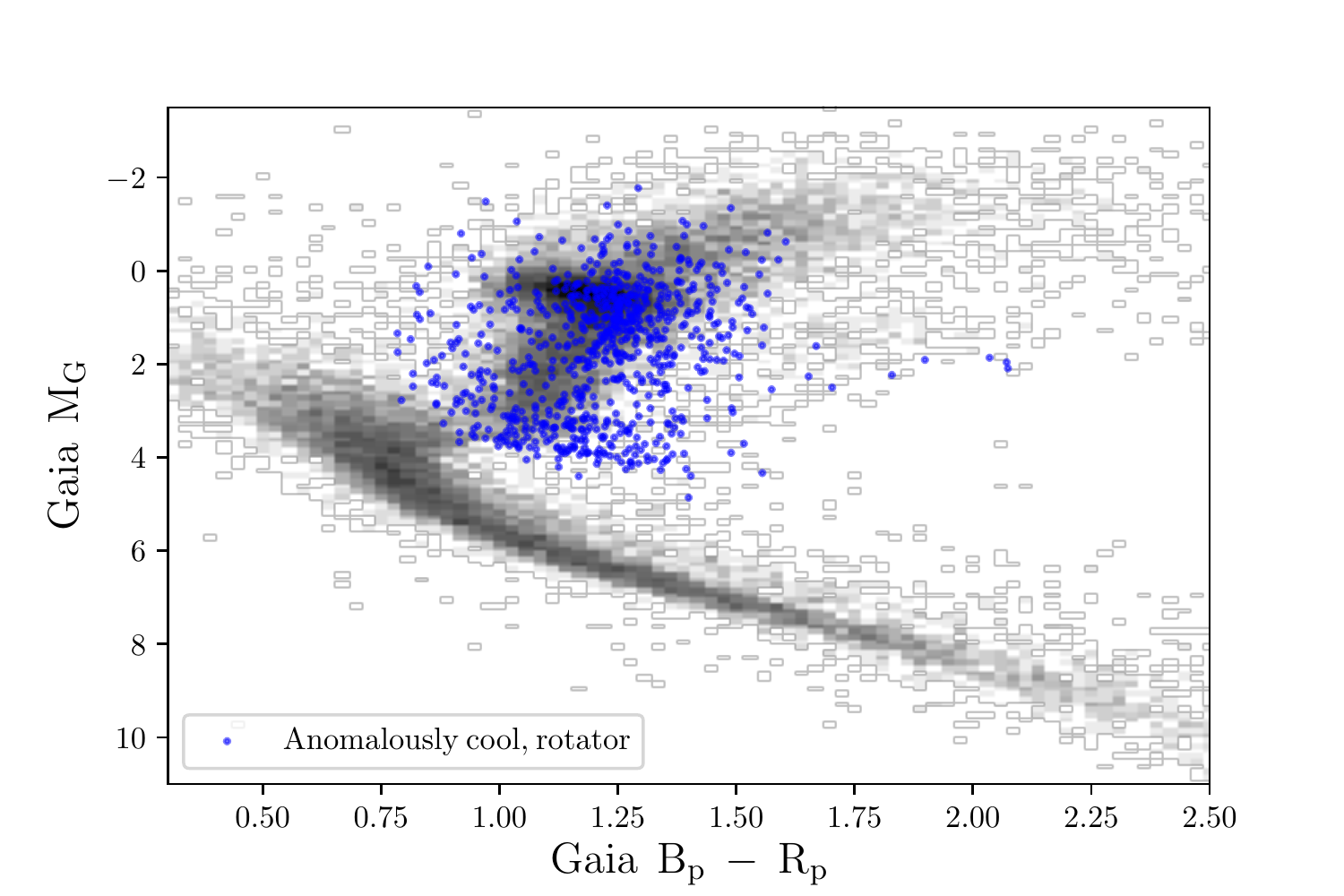}
    \includegraphics[scale=0.56,trim={0.0cm 0.1cm 0.5cm 0.5cm},clip]{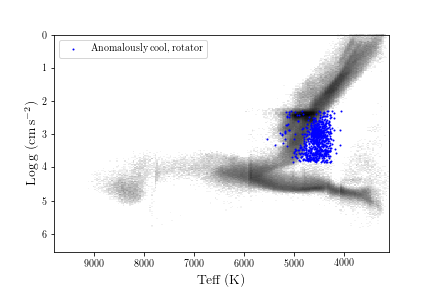}
    \includegraphics[scale=0.56,trim={0.0cm 0.1cm 0.5cm 0.5cm},clip]{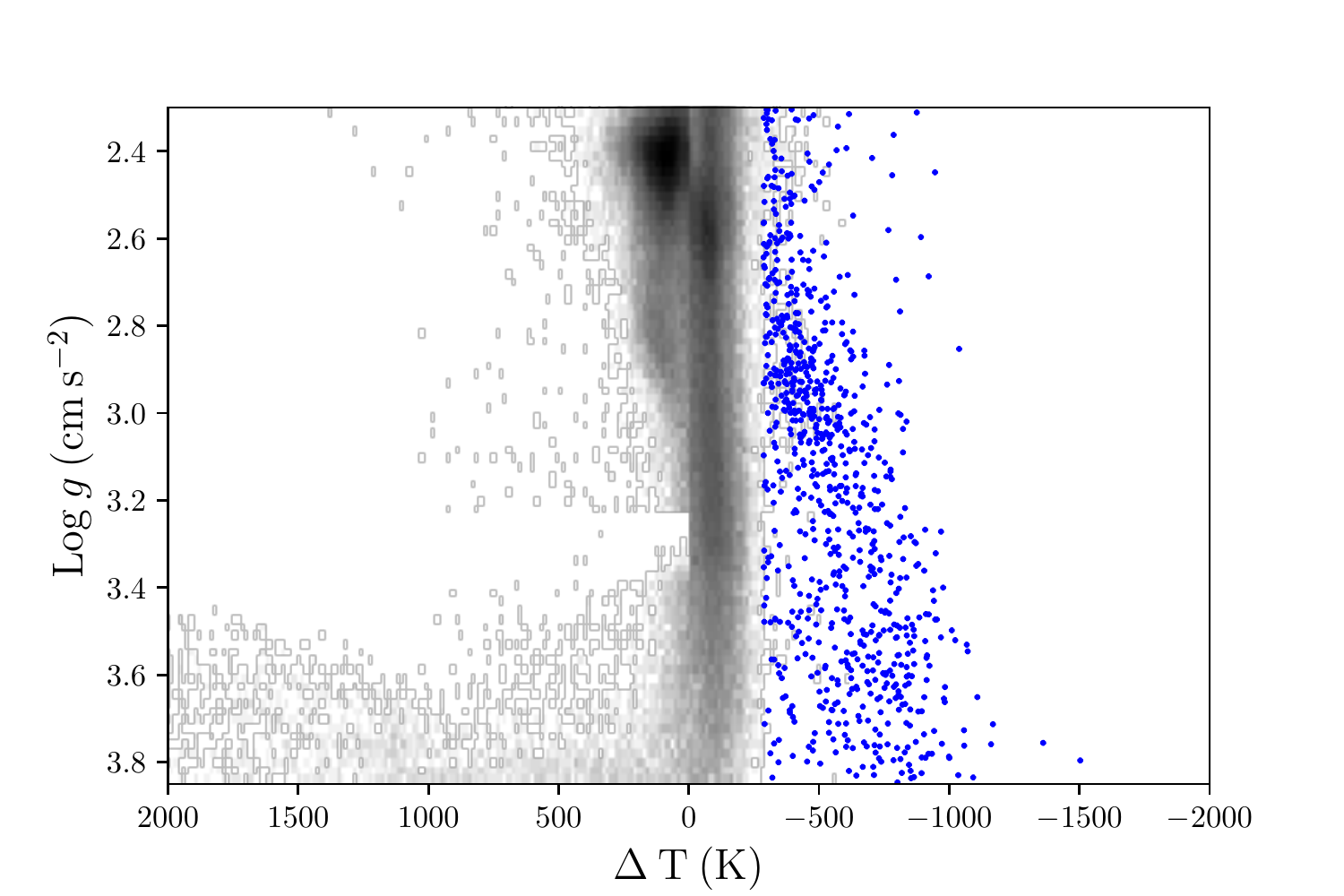}
    \caption{The distribution in a \textit{Gaia} CMD (top), log \textit{g} versus $T_\mathrm{eff}$ (middle), and log \textit{g} versus $\Delta T$ (bottom) of the confirmed anomalously cool rapid rotators. The grey histogram represents a subset of the entire DR16 sample.}
    \lFig{dr16_rot}
\end{figure}

\begin{figure}
    \centering
    \includegraphics[scale=0.56,trim={0.0cm 0.1cm 0.5cm 0.5cm},clip]{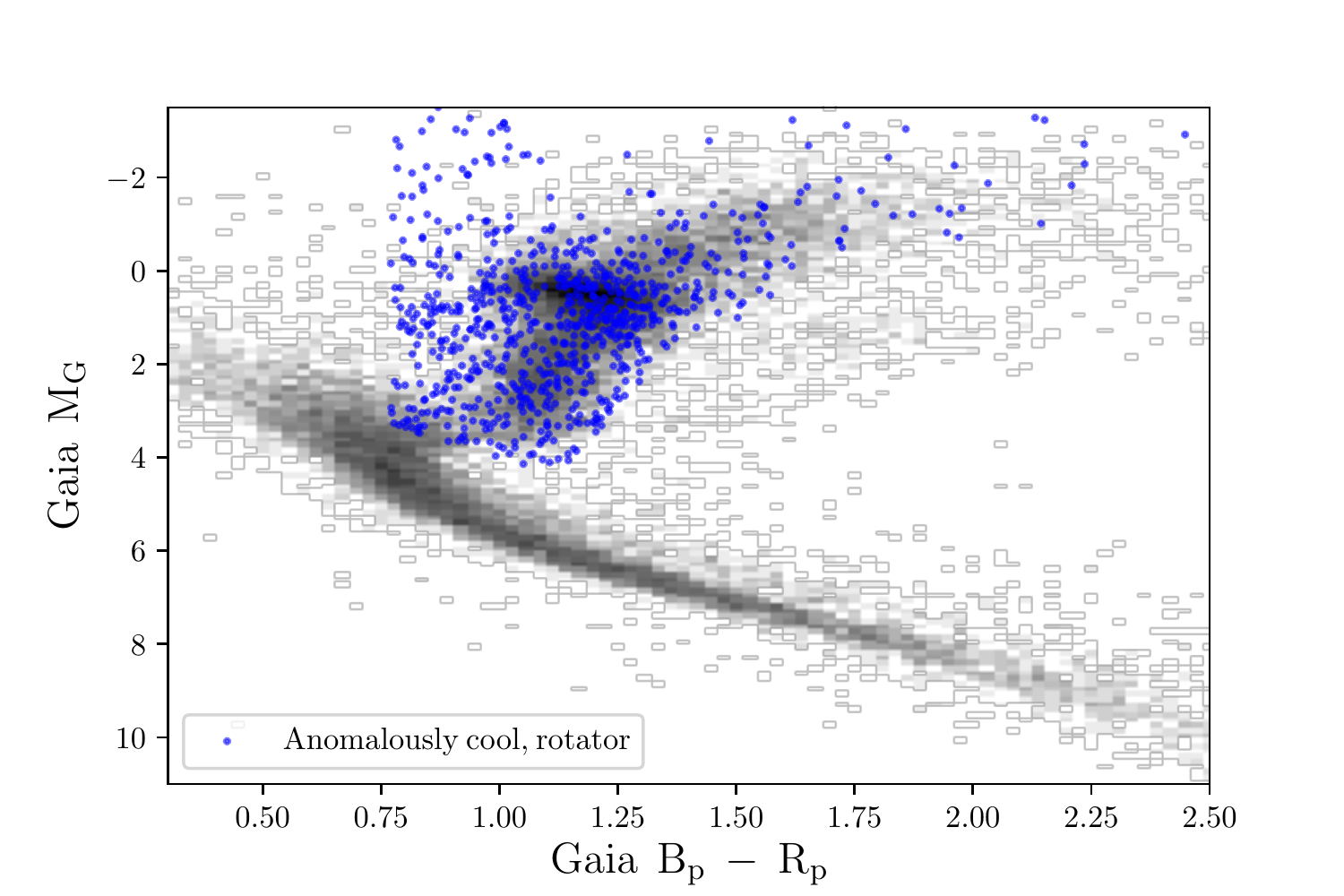}
    \caption{A \textit{Gaia}  CMD of the confirmed rapid rotators with partial spectroscopic solutions. The grey histogram represents a subset of the entire DR16 sample.}
    \lFig{partial_rap_CMD}
\end{figure}

\begin{figure}
    \centering
    \includegraphics[scale=0.55,trim={0.0cm 0.1cm 0.5cm 0.5cm},clip]{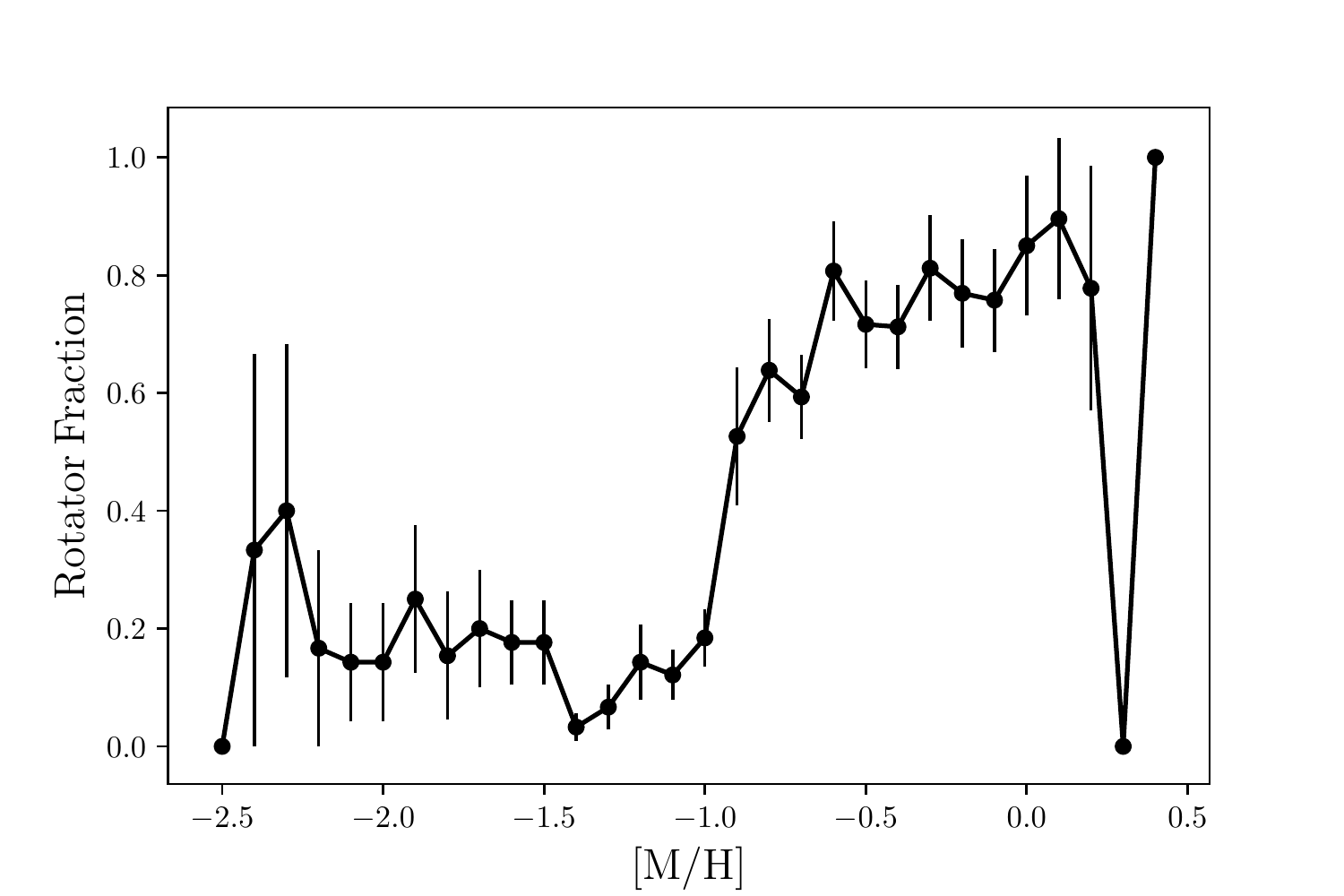}
    \caption{The fraction of giants in a bin of 0.1 dex in [M/H] that are rapidly rotating. The errors bars represent Poisson noise. There is an apparent metallicity dependence, with a break at [M/H] = -1, where we see that few of the low metallicity stars are rapid rotators.}
    \lFig{rap_frac}
\end{figure}

Because the field stars in DR16 are less well-characterized than those in the \textit{Kepler} field, we take a more careful approach when identifying the rapid rotators by extracting \textit{v}sin\textit{i}. The ASPCAP \citep{Gar16} pipeline has been applied successfully in the characterization of large samples of giants; however, it does not currently have a prescription to measure rotational velocity in spectra simultaneously with other stellar parameters. To obtain rotational velocity estimates from APOGEE H-band spectra we modify the existing ASPCAP pipeline, which includes seven dimensions including microturbulence, [C/M], [N/M], [alpha/M], [M/H], surface gravity, and effective temperature, with a fit for macroturbulence. We use a fiducial value of 2 km/s in microturbulence and replace this dimension with surface rotation (sampled at $v \, \sin \, i$ of 1.5, 3.0, 6.0, 12.0, 24.0, 48.0, and 96.0 $km/s$) in our modified 7-D fit. We use stellar atmospheres from MARCS with the same extent in metallicity, chemical abundances, and effective temperatures as in the ASPCAP pipeline; we kept the grid spacing similar in most dimensions but reduced it from 0.25 to 0.5 dex in others for computational purposes, which does not significantly impact stars with S/N > 70 \citep{Gar16}. Combined spectra from APOGEE DR17 were continuum normalized simultaneously with the synthetic spectra and fit with {\sc{ferre}}\footnote{{\sc ferre} is publicly available from \url{http://hebe.as.utexas.edu/ferre}.}, a least-squares fitting spectra analysis code. Stellar parameters were obtained through the unconstrained optimization by quadratic approximation algorithm and errors were estimated through inversion of the curvature matrix \citep{Gar16}. 

We compare our measured log \textit{v}sin\textit{i} from our sample to a subset of 1021 regular APOGEE red giants on the lower RGB in \Fig{lyra}. Two things are apparent. The first is that the regular stars rarely exceed a log \textit{v}sin\textit{i} of 0.7 km s$^{-1}$; they are true slow rotators as expected. The second is that there is a break in the distribution of log \textit{v}sin\textit{i} at 0.7 instead of 1, as we would expect if the threshold for rapid rotation were 10 km s$^{-1}$. Instead, this gap suggests that the distinct population of rapid rotators starts at 5 km s$^{-1}$, consistent with \citet{Tay15}. As such, we adopt log \textit{v}sin\textit{i} of 0.7 km $s^{-1}$ as our threshold for rapid rotation in the \textit{v}sin\textit{i} we measure ourselves. For the values reported by APOGEE, we maintain the 10 km s$^{-1}$ threshold. 

Using \textit{v}sin\textit{i} from APOGEE, our measured \textit{v}sin\textit{i}, and \textit{Gaia} $v_\mathrm{Broad}$, we confirm rotation in 1598 (50\%) stars, with 1188 having \textit{v}sin\textit{i} > 10 km s$^{-1}$ and 410 having 5 < \textit{v}sin\textit{i} < 10 km s$^{-1}$. Given that we have no constraint on the inclinations of these systems, some stars will be true rapid rotators with low \textit{v}sin\textit{i}. For example, 15\% of stars with random inclinations have sin\textit{i} < 0.5. Rapid rotation could still induce spectroscopic anomalies even if the star is viewed pole-on - a large star spot filling fraction from stellar activity, for example. Correcting for random inclination effects, our data corresponds to a true rapid rotator fraction of 65\%. 

Like our test sample from the APOKASC-3 catalog, the vast majority of these stars have \textit{v}sin\textit{i} > 10 km s$^{-1}$, further supporting that, at least in DR16, significant rotation is needed in order to induce spectroscopic anomalies. Based on the yield from the anomalously cool stars in APOKASC-3, we expect 979 anomalously cool giants to have \textit{v}sin\textit{i} > 10 km s$^{-1}$. In DR16, we find 541 (55\%). Furthermore, in APOKASC-3, both criteria recover 56\% of the total number of giants rotating with at least 10 km s$^{-1}$. Based on this yield, our catalog captures 1188 (66\%) of the expected 1802 giants with \textit{v}sin\textit{i} > 10 km s$^{-1}$. 

\Fig{dr16_rot} shows the observed properties of the rapid rotators. The anomalously cool rotators do not favor a particular range in log \textit{g} nor do they cluster around a particular $\Delta T_\mathrm{eff}$. Like the APOKASC-3 sample, most rotators are cooler than the giants, but not entirely separated from the giant branch. The same is true in the \textit{Gaia} CMD. The rapid rotators fall on the lower giant branch, though their median color is 0.13 magnitudes redder than the non-rotating, metal-rich stars. This could be due to spots. The rapidly rotating giants with partial solutions also fall predominantly on the lower giant branch, as shown in \Fig{partial_rap_CMD}, completely indistinguishable from typical giants. 

We know from the APOKASC-3 sample that our spectroscopic criteria preferentially identify rotators with \textit{v}sin\textit{i} > 10 km s$^{-1}$ and that the number of rotators with 5 < \textit{v}sin\textit{i} < 10 km s$^{-1}$ far outnumbers those with \textit{v}sin\textit{i} > 10 km s$^{-1}$. While this could be a population effect, it is important to see if the same pattern is observed in the field. We use the same control sample as above, randomly selected from lower giant branch stars (with 2.3 < log \textit{g} < 3.6 cm s$^{-2}$) that we photometrically confirm as giants. Among these 1021 giants, we recover the same general trends of rapid rotation as the APOKASC-3 sample, even when the APOKASC-3 sample is restricted to the same log g, though at half the same absolute rate. 2\% (18/1021) of the DR16 control sample rotates with 5 < \textit{v}sin\textit{i} < 10 km $^{-1}$, compared to 4\% in the APOKASC-3 sample, and 0.4\% (4/1021) of the DR16 control rotatoes with \textit{v}sin\textit{i} > 10 km s$^{-1}$, compared to 1\% in the APOKASC-3 sample. 

There is also a striking trend with metallicity, but not in the direction we might expect. \Fig{rap_frac} shows that the fraction of anomalously cool giants where we confirm rapid rotation is half to a quarter of the rates for higher metallicity stars. There are several things at play here. The first is that there is a metallicity offset of -0.37 dex between the median [M/H] of all anomalously cool giants and a random sample of ``regular" giants. This suggests that in addition to a lower $T_\mathrm{eff}$ and a higher log \textit{g}, ASPCAP also returns lower [M/H]. 

The second is that the anomalously cool criterion catches more metal-poor stars than metal-rich stars. \Fig{dr16_cool_abun} shows that the scatter in $\Delta T$ below [M/H] of -1 is much larger. As such, the anomalously cool criterion captures 18\% of all metal-poor giants compared to just 1.5\% of metal-rich giants. If we correct for the -0.37 dex metallicity offset, the fraction of anomalously cool, metal-poor giants drops to 9\%. This implies two things. The first is that 10\% of metal-poor giants are false positives; it is a result of a poor fit that the metallicity is low. The second is that the temperature errors on the low metallicity stars are larger than for those with high metallicity. This is reflected in the distribution of $\Delta T$ values shown in \Fig{dr16_cool_abun}. The anomalously cool criterion is defined to be a 3$\sigma$ deviation from the median values of each metallicity bin. However, when we fit a Gaussian distribution to the $\Delta T$ values of giants in each metallicity bin below -1, we see non-Gaussian tails and large scatter such that the 3$\sigma$ boundary to separate the anomalously cool giants is on average only 1.14 standard deviations from the peak. As such, if there are large temperature errors in for these metal-poor giants, $Delta T$ is larger and we mistakenly flag more of the population as anomalously cool. Furthermore, since there is such a broad scatter in $\Delta T$ among the rapid rotators and the false positives, there is no way we could have defined a boundary where rapid rotators are preferentially identified.

The third is that there are also just fewer rapid rotators at lower metallicity. The drop in the rapid rotator fraction at low metallicity results from the combination of a fewer rapid rotators and a higher fraction of the overall metal-poor giant population being deemed anomalously cool. The numerator is small and the denominator is large compared to the metal-rich giants. 

\subsection{The origin of rapid rotation}

 \begin{table}
    \caption{Four different measures of binary fractions of the confirmed rapid rotators in DR16.} 
    \centering
    \begin{tabular}{lcc}
         &  \textit{v}sin\textit{i} > 10 km s$^{-1}$ & 5 < \textit{v}sin\textit{i} < 10 km s$^{-1}$\\
         \hline
       Anomalously cool & 541 & 281\\
       RUWE f$_\mathrm{bin}$  & 16$\pm$2\% (86/537) & 21$\pm$3\% (59/277)\\
       \textit{Gaia} RV f$_\mathrm{bin}$ & 78$\pm$7\% (138/178) & 70$\pm$9\% (66/94)\\
       APOGEE f$_\mathrm{bin}$ & 88$\pm$4\% (381/434) & 88$\pm$6\% (207/234)\\
       APOGEE f$_\mathrm{bin,close}$ & 74$\pm$4\% (322/434) & 79$\pm$6\% (185/234)\\
       \hline
       Partial solution & 647 & 129\\
       RUWE f$_\mathrm{bin}$  & 19$\pm$2\% (124/644) & 26$\pm$4\% (33/128)\\
       \textit{Gaia} RV f$_\mathrm{bin}$ & 74$\pm$5\% (206/279) & 52$\pm$14\% (13/25)\\
       APOGEE f$_\mathrm{bin}$ & 70$\pm$4\% (352/506) & 50$\pm$8\% (38/76)\\
       APOGEE f$_\mathrm{bin,close}$ & 54$\pm$3\% (274/506) & 28$\pm$6\% (21/76)\\
       \hline
    \end{tabular}
\end{table}

To begin to understand the origin of rapid rotation in our sample, we search for binary companions among our rapid rotators. We see the same general trends as in the APOKASC-3 sample. The anomalously cool giants have a higher binary fraction than the giants with partial solutions and the majority of the anomalously cool binaries are close. However, the giants with partial solutions show lower binary fractions across all measures. We also find that about 20\% of stars have a wide binary companion, comparable to what we find in APOKASC-3 but still higher than anticipated. We discuss this further in \Sect{disc}. Table 8 lists the binary fractions of this sample.
 
 We suspect that the remaining giants which had enough data to assess binarity but had no indication of a companion are merger products. They occupy the same space as the binary rapid rotators in a \textit{Gaia} CMD, and those with complete spectroscopic solutions are indistinguishable from binaries in log \textit{g} - $T_\mathrm{eff}$ space. Among the stars with complete spectroscopic solutions the fraction of single rapid rotators is strongly metallicity-dependent. Below [M/H] of -1, 36\% of rapid rotators are single, compared to 7\% at high metallicity, after correcting for the metallicity bias. This is not a byproduct of ASPCAP, as \citet{Car11} also found that their population of single rapidly rotating K giants had lower [Fe/H] metallicity (0.6 dex on average) than their slowly rotating counterparts. Several things could lead to this. These giants could be merger products, either with a stellar or substellar companion. Since the close binary fraction is higher at low metallicity and there i evidence that the period distribution of those bianries is also metallicity-dependent \citep{Moe19}, a merger scenario makes sense. Low metallicity stars are also subject to less mass loss, which would weaken magnetic breaking on the main sequence and keep the star spinning more rapidly as it evolved to the giant branch. Also, we are subject to small number statistics here as there are only 11 single rapid rotators with [M/H] < -1. 

\section{Discussion}
\lSect{disc}

Individual samples of rapidly rotating red giants typically contain on the order of tens to hundreds of stars \citep[e.g.,][]{Tay15, Cei17, Gau20}. With APOGEE, we can extract samples an order of magnitude larger \citep[e.g.,][]{Dah22}. With our new spectroscopic criteria, we have created an sample of 1598 rapid rotators. We also provide a comparably-sized cohort of stars (3376 targets) without spectroscopic solutions. With independent validation, this group would yield a significant increase in the rapidly rotating population. While the stars in this catalog are interesting in their own right, we have also uncovered many unforeseen but interesting results during the catalog assembly process. 

The first is that the population of rapidly rotating red giants is far larger than we previously thought. With approximately 10\% of giants expected to interact at some point on the RGB \citep{Car11, Tay15,Bad18}, finding that 4.9\% of APOKASC-3 giants and 2.5\% of DR16 giants rotate with at least 5 km s$^{-1}$ implies that the number of true rapid rotators in APOGEE is in the tens of thousands. This is seen in \citet{Dah22} as well. No one criterion will capture all of these rapid rotators, especially since we are not sensitive to the majority of them. By definition, the anomalously cool criterion is only sensitive to the lower RGB. While giants with partial solutions span the giant branch, they make up a minority of giants in all of DR16. Furthermore, the bulk of these rapid rotators, especially those at the 5 km s$^{-1}$ threshold do not even appear to be spectroscopically anomalous, and there are $\sim$ 4x as many stars rotating at the 5 km s$^{-1}$ threshold compared to the 10. Without measuring their \textit{v}sin\textit{i}, we would not know that they are rapid rotators.

These stars can only be identified if rotation is included as a free parameter in the giant grids used in ASPCAP. This is a computationally expensive task, however, there are two strong reasons to do this. The first, is that there are tens of thousands of post-interaction red giants in the field that are currently undetected. Finding these giants completes the census of interacting binaries, allowing us to better constrain the relative rates and outcomes of different types of binary interaction. The second is that including rotation in our independent spectroscopic fits allowed us to obtain fits for giants where ASPCAP could not find solutions, and it improved the fits for stars where ASPCAP could find solutions. We saw shifts in the stellar parameters returned by ASPCAP for the anomalously cool stars to more realistic values, such as increase in [M/H] and decreases in log \textit{g}. Including rotation for the giants not only enables more accurate fits, it also enables more fits in the first place.

Furthermore, measuring \textit{v}sin\textit{i} for all giants ensures a uniform assessment of rotation. Aggregating rapid rotators from different surveys, or searching for rotators using a variety of techniques, leads to complicated selection functions and biases that are difficult to correct for. For example, identification of rapid rotators from rotational modulation is biased towards higher spot amplitudes, and the telescope with which a star is observed leads to a bias in rotation periods; TESS is biased to short periods whereas \textit{Kepler} can detect longer periods. A lack of uniform measures of rotation make it difficult to assess completeness.

The biases might also not be what one might think. This leads to the second surprising result. We thought that our spectroscopic selection criteria, especially the anomalously cool stars, would almost exclusively pick out tidally-synchronized binaries. They instead yielded a mix of systems. While we found that the majority of giants in our sample are rapid rotators in binaries, the close binary fraction for these giants was lower overall than anticipated. This is the opposite of what one would expect. The fraction of longer period binaries, probed by RUWE and the less strict APOGEE RV criterion, are also non-negligible. Rapid rotation in a giant with a distant companion is difficult to explain, since there is no interaction to cause spin-up. However, three-body interactions could explain these systems if instead, the distant companion was actually a tertiary and the rapid rotator resulted from a merger of the inner binary. It is outside the scope of this paper to investigate the feasibility and expected rates of three-body interactions producing these systems. We present it here as a possible solution. 

While our criteria are effective at finding rapid rotators, we have learned that we cannot use these criteria to preferentially identify rapid rotators of a singular origin. To pick out just the tidally synchronized systems, the contact binaries, or the merger products, for example, one would need a \textit{v}sin\textit{i} measurement as well as a careful selection function to isolate stars in particular regions of the HR diagram or spectroscopic stellar parameter space. 

However, using \textit{v}sin\textit{i} to identify rapid rotators has its shortcomings too. Fixed values of \textit{v}sin\textit{i} do not correspond to the same rotation periods due to the factor of 10 change in radius between the lower and upper RGB. The convective overturn timescales also increase as a star moves up the giant branch meaning that comparable \textit{v}sin\textit{i} values in stars do not necessarily imply comparable levels of activity. Furthermore, ASPCAP can return spuriously high \textit{v}sin\textit{i} measurements for more slowly rotating but equal mass (comparable luminosity) binary stars due to the confusion between orbital and rotation period \citep{Sim19}. While we do not expect this to be a significant source of contamination, we still must be aware of it.

Nevertheless, this catalog is a unique grouping of rapid rotators whose rapid rotation originates from a variety of sources. There are RS CVn variables, merger products, and synchronized binaries. The catalog could be searched for cataclysmic variables or non-interacting compact object companions. The merger and binary populations taken individually can probe a variety of interesting astrophysics. For example, a comparison between the number of mergers seen versus how many we would expect to see based on the time it takes to spin down post-merger can calibrate spin-down models. The properties of the binary systems, such as their mass and period distributions, can be compared to known binary distributions, like the ones in \citet{Moe17}. For stars with seismology, the internal structure of interacting or post-interaction giants can be assessed, or, masses derived from binary systems can calibrate asteroseismic scaling relations (e.g., Beck et al., in prep). While assembling this catalog has revealed more about the underlying population of rapid rotators, the catalog itself is a useful resource for studying en masse giants with unique evolutionary histories.

Beyond what can be done with this catalog, is the question of whether the same or similar criteria can be applied to subsequent APOGEE data releases or other large spectroscopic surveys. In principle they can, but how well the criteria perform depends on the quality of the spectra and the performance of the processing pipelines. 

In APOGEE DR17 \citep[][Holtzman et al. in prep]{Abd22}, the anomalously cool criterion is still valid, but it is not as good at capturing rapidly rotating red giants. The $T_\mathrm{ref}$ calculation for DR17 is
\begin{equation}
    T^{'}_\mathrm{ref} = 3082.884 -533.781 \mathrm{[Fe/H]} + 553.171 \mathrm{log} g - 282.087\mathrm{[C/N]},
\end{equation}
using the stellar parameters derived in DR17. When we take our sample of 178 APOKASC-3 targets and see if we can identify them as spectroscopic outliers using DR17 parameters, we cannot recapture the sample. 37 stars have log \textit{g} and $T_\mathrm{eff}$ outside the range where the anomalously cool criterion is valid. Of the 95 stars with stellar parameters in the range where we can calculate temperature offsets, about half appear anomalously cool. The last 46 stars have no spectroscopic solution, but 35 of them were confirmed  rapid rotators. In DR17, selecting giants with no spectroscopic solution could yield a high-fidelity sample of rapid rotators. We cannot study stars with partial solutions in DR17 because these types of stars do not exist in DR17. All stars that had initial FERRE fits were pushed through the pipeline; they had complete fits.A combination of giants with no spectroscopic solution and anomalously cool giants should recover roughly one third of the expected population of rapid rotators in DR17. In DR17 ASPCAP did a better job of getting more accurate stellar parameters for rapidly rotating stars.

Regardless of the criterion adopted to find rapidly rotating red giants in a particular spectroscopic survey, it is promising that this analysis is possible at all. Being able to exploit lower quality data and systematic errors in a pipeline to find astrophysically interesting stars helps maximize the utility of the survey. 

The existence of lower quality data also means that there are gains to be made in processing the spectra. We are able to fit all but 29 of them. Allowing rotation to be a free parameter changes the other stellar parameters as well. The median log \textit{g} in our sample is higher by 0.13 dex and the median [Fe/H] is lower by 0.25 dex compared to our independent spectral fits. The independent fits shifted the median $T_\mathrm{eff}$ colder by about 150 K, opposite of the direction we expect. However, this may be explained by the presence of star spots. While a spot analysis \citep{Cao22} is not included in this work, we expect many of these stars to be active, and the presence of spots could further bring down $T_\mathrm{eff}$. 

Moving forward, spectroscopic identification through \textit{v}sin\textit{i} measurements or photometric detection through spot modulation seem to be the most promising means for identifying rapid rotators. The disagreement between $v_\mathrm{Broad}$ and \textit{v}sin\textit{i} makes \textit{Gaia} $v_\mathrm{Broad}$ difficult to use. Furthermore, less than 500 giants in our sample had \textit{Gaia} $v_\mathrm{Broad}$ measurements. Rotation periods from TESS offer a promising means to find rapid rotators, especially with a new pipeline which can extract rotation periods longer than the 13.7 day limit imposed by TESS's 27 day viewing window \citep{Cla22}. 

\section{Conclusions}
\lSect{con}
For giants born below the Kraft break, rapid rotation is a direct consequence of binary interaction. These products of interaction produce sets of astrophysically interesting systems which follow evolutionary tracks often not accessible in single star evolution, including evolved blue stragglers, merger products, active giants, and contact binaries, to name a few. In this paper, we have identified a large cohort of rapid rotator candidates selected from evolved, cool giants in APOGEE. 

We find that the three spectroscopic categories of giants with no solution, giants with partial solutions, and giants with complete solutions, created as APOGEE assesses the quality of its spectra, make natural and efficient groups for identifying rapid rotators. Anomalously cool giants with complete solutions and giants with partial solutions are particularly good at preferentially identifying rapid rotators, with rotator fractions of 90\% and 78\% respectively in our APOKASC-3 control sample (\Fig{cool}, \Fig{partial}). The majority of these stars rotate with \textit{v}sin\textit{i} > 10 km s$^{-1}$. This makes sense since we are targeting the biggest spectroscopic outliers and we know that rapid rotation can change reported stellar parameters like log \textit{g} \citep{Tho19} and $T_\mathrm{eff}$ \citep{Dix20}. The rotator fraction at the >5 km s$^{-1}$ threshold drops to 55\% and 45\% in our DR16 catalog, which, while still high, could result from population differences between the \textit{Kepler} field, which is old and above the plane, and general field stars. The difference could also result from the different ways we measure \textit{v}sin\textit{i} in the two populations, though we detect no systematic biases in our \textit{v}sin\textit{i} measurements in the APOKASC-3 field.

We also discovered two failure modes of ASPCAP. In both the APOKASC-3 and DR16 samples 15\% of stars identified as anomalously cool giants were actually main sequence photometric binaries based on their \textit{Gaia} photometry (\Fig{cool} and \Fig{dr16_cool}). For stars where both dwarf and giant template spectra were fit, the giant templates fit these spectra slightly better. For stars at the base of the giant branch we recommend photometric verification of stars selected to be spectroscopic giants. 

The second failure mode is a metallicity bias. We found two issues with this low metallicity population. The first is that the median [M/H] for the anomalously cool giants is 0.37 dex lower than the median for ``normal" red giants. As a result, intrinsically higher metallicity stars, which are more common, scatter into the poorly populated metal-poor domain, artificially inflating the apparent fraction of metal-poor rotators. Correcting for this offset resulted in a 10\% reduction in the total number of low-metallicity giants. The second is that large random errors in temperature for metal-poor stars make our spectroscopic temperature anomaly an inffective diagnostic of rapid rotation there, resulting in a large number of false positives. The net result is that our method does not yield strong results about rapid rotation in the metal-poor domain.

In assessing the completeness of our spectroscopic selection criteria we measure \textit{v}sin\textit{i} for all of the giants in APOKASC-3. This revealed a population of giants rotating with 5 < \textit{v}sin\textit{i} < 10 km s$^{-1}$ that was almost 4x larger than those with \textit{v}sin\textit{i} > 10 km s$^{-1}$. This is not a population effect. In a control sample of ``normal" DR16 giants we find the same relative rates of rapid rotation at the 5 and 10 km s$^{-1}$ threshold. Our criteria are largely insensitive to these stars (\Fig{Api}), meaning that giants which are rotationally enhanced fly under the radar due to their lack of spectroscopic anomalies. 

We therefore recommend that current and future spectroscopic surveys include rotation as a free parameter in their solutions for cool evolved stars. It is a computationally expensive task, but the payoff is a census of interacting and post-interaction binaries on the order of ten thousand. \textit{v}sin\textit{i} is the only way to have a uniform measure of rotation, as we cannot guarantee rotation periods from TESS for all of these targets. \textit{Gaia} is not helpful here either because there are $v_\mathrm{Broad}$ measurements for only some DR16 giants and it is only sensitive down to 10 km s$^{-1}$. Furthermore, our criteria are mostly sensitive to rapid rotation on the lower RGB. It is only with \textit{v}sin\textit{i} measurements for all giants that we can find the rotators on the upper RGB and in the clump. 

Lastly, \textit{v}sin\textit{i} measurements for all giants lets us first identify the rapid rotators and then assess the binary origins of their rotation. This is important because our spectroscopic selection criteria do not actually isolate the systems we thought they would. We expected to almost exclusively capture tidally-synchronized binaries on the lower RGB, but the close binary fraction was only about two-thirds in the APOKASC-3 sample. This means that our criteria identify giants with a variety of binary origins. To select rapid rotators that have particular features, like those which are just tidally-synchronized, just RS CVns, or just merger products, more careful consideration must be given to where those systems would cluster in a CMD or in spectroscopic parameter space. 

Rapid rotation in giants is more common than we expected. We have shown this by producing the largest catalog of both single and binary rapidly rotating red giants that are not exclusively variable to date. We have shown how to exploit pipeline failure modes to find astrophysically interesting systems quickly and with high fidelity, and we have shown that the prevalence of binary interaction in low-mass giants is much greater than previously thought. Measurements of \textit{v}sin\textit{i} are good ways to find interacting and post-interaction binaries. With subsequent APOGEE data releases and the wealth of stellar spectra from SDSS-V, we are poised to fill in gaps in the census of rapidly rotating red giants if rotation is included for all giants when fitting spectra. An accurate census and careful characterization of these systems are critical for concretely testing predictions from binary population synthesis, probing mass transfer physics and common envelope evolution, and generally exploring the effects of binary interaction on evolved stars. 

\section*{Acknowledgements}
RAP would like to thank Diego Godoy-Rivera for providing the de-reddening scripts and for guidance on how to process the \textit{Gaia} photometry. S.M.~acknowledges support from the Spanish Ministry of Science and Innovation (MICINN) with the Ram\'on y Cajal fellowship no.~RYC-2015-17697, grant no.~PID2019-107187GB-I00 and PID2019-107061GB-C66, and through AEI under the Severo Ochoa Centres of Excellence Programme 2020--2023 (CEX2019-000920-S).

\section*{Data Availability}
The catalog of rapidly rotating stars we identify in APOKASC-3 and APOGEE DR16, as well as the stellar parameters we use to identify them, are all available upon request. Upon acceptance to the journal, the data will be made publicly available in tables released with this paper. 

\bsp	
\label{lastpage}
\end{document}